\documentclass[12pt,preprint]{aastex}
\usepackage{amsmath}
\usepackage{amssymb}
\usepackage{graphicx}
\usepackage{subfig}

\slugcomment{September 16, 2009}

\def\bfnabla{{\mbox{\boldmath $\nabla$}}}
\def\kms{\,\mbox{km s}^{-1}}
\def\kpc{\,\mbox{kpc}}
\def\pc{\,\mbox{pc}}
\def\yr{\,\mbox{yr}}
\def\Gyr{\,\mbox{Gyr}}
\def\au{\,\mbox{AU}}
\def\msun{M_\odot}

\shortauthors{Y.-F. Jiang \& S.\ Tremaine} \shorttitle{Binary Stars}

\begin{document}

\title{The Evolution of Wide Binary Stars}

\author{Yan-Fei  Jiang\altaffilmark{1}  \& Scott Tremaine\altaffilmark{2}  }
\affil{$^1$Department of Astrophysical Sciences, Princeton
University, Princeton, NJ 08544, USA} \affil{$^2$ School of Natural
Sciences, Institute for Advanced Study, Einstein Drive, Princeton,
NJ 08540, USA}

\begin{abstract}
  We study the orbital evolution of wide binary stars in the solar
  neighborhood due to gravitational perturbations from passing stars. We
  include the effects of the Galactic tidal field and continue to follow the
  stars after they become unbound. For a wide variety of initial semi-major
  axes and formation times, we find that the number density (stars per unit
  logarithmic interval in projected separation) exhibits a minimum at a few
  times the Jacobi radius $r_J$, which equals $1.7\pc$ for a binary of
  solar-mass stars. The density
  peak interior to this minimum arises from the primordial distribution of
  bound binaries, and the exterior density, which peaks at $\sim
  100$--$300\pc$ separation, arises from formerly bound binaries that are
  slowly drifting apart. The exterior peak gives rise to a significant
  long-range correlation in the positions and velocities of disk stars that
  should be detectable in large astrometric surveys such as GAIA that can
  measure accurate three-dimensional distances and velocities.

\end{abstract}


\section{Introduction}

\label{sec:intro}

Wide binary stars are disrupted by gravitational encounters with
passing stars, molecular clouds, and other perturbers. This process
was first investigated by \cite{opik1932}, who estimated the
$e$-folding time for disruption of binaries composed of solar-mass
stars, with apocenter distances of $1\pc$, to be $10\Gyr$ or less.
Other early estimates of the disruption rate are due to
\cite{ambart1937}, \cite{chandra1944}, \cite{yabu1966},
\cite{heggie1975}, \cite{king1977}, \cite{heggie1977},
\cite{rk1982}, and \cite{bahcalletal1985}. In the last of these a
binary of age $t_0$ and component masses $M_1$ and $M_2$ was
estimated to have a 50\% survival probability at semi-major axis
\begin{equation}
a_{1/2}(t_0)=0.002\frac{(M_1+M_2)\sigma}{G\rho_2t_0}=3.1\times
10^4\au\, \frac{M_1+M_2}{2\msun}
\frac{\sigma}{50\kms}\frac{0.03\msun^2\pc^{-3}}{\rho_2}\frac{10\Gyr}{t_0}.
\end{equation}
Here $3\sigma^2$ is the mean-square relative velocity between the
center of mass of the binary and the perturbing stars, and
$\rho_2\equiv\int n(M_p)M_p^2dM_p$ is the second moment over mass of
the number density of stars in the solar neighborhood (cf.\ eq.\
\ref{massdensity}). This estimate is substantially shorter than
\"Opik's, mostly because it includes the cumulative effects of
distant, weak encounters (which \"Opik recognized to be important
but did not compute).

A closely related problem is to estimate the distribution of
semi-major axes of wide binaries. For relatively small semi-major
axes $a\lesssim a_{1/2}(t_0)$ the distribution is presumably
primordial, and thus reflects the (poorly understood) formation
process of wide binaries. At larger semi-major axes $a\gtrsim
a_{1/2}(t_0)$ the distribution ought to be primarily determined by
the disruption process. The Fokker--Planck equation that describes
the evolution of the semi-major axis distribution (eq.\
\ref{diffCBS}) was derived and solved by \cite{king1977},
\cite{rk1982} and \cite{Weinbergetal1987}, who showed that
$dn\propto da/a^2$ for $a\gtrsim a_{1/2}(t_0)$.

Observationally, the distribution of wide binary semi-major axes is
determined by measuring the projected separations of
common-proper-motion binaries (e.g., \citealt{chanamegould2004},
\citealt{pov07}, \citealt{lepinebongiorno2007},
\citealt{sesaretal2008}; see also \citealt{chaname2007} and
references therein). For $a\lesssim 3\times10^3\au$ the distribution
of separations or semi-major axes\footnote{For a population of
binaries at a given semi-major
  axis $a$, with other orbital elements assigned as described at the start of
  \S\ref{simulation}, the median projected separation is $0.978a$.  Thus we
  may assume that the distributions of semi-major axes and separations are nearly
  the same.} of disk binaries is approximated well by \"Opik's (1924) law,
\begin{equation}
dn \propto d\log a = \frac{da}{a}. \label{eq:opikinit}
\end{equation}
At larger semi-major axes, the number of binaries falls more
steeply, roughly as $dn\propto da/a^{1.6}$ for $3\times10^3\au
\lesssim a \lesssim 10^5\au$ \citep{lepinebongiorno2007}. The
further steepening to $dn \propto da/a^2$ that is expected for
$a\gtrsim a_{1/2}(10\Gyr)\simeq 3\times 10^4\au$ is much more
difficult to detect. There have been a number of claimed detections
of this steepening---often, less accurately, called a
``cutoff''---but these are controversial
\citep{bs1981,ww1987,latham1991,ww1991,pal2000,yooetal2004,quinnetal2009}.
Measurements of the semi-major axis distribution are likely to
improve dramatically in the next few years because of large,
accurate proper-motion surveys. In particular, the GAIA spacecraft
will determine both proper motions and trigonometric parallaxes for
millions of nearby stars with unprecedented accuracy, allowing a far
better determination of the binary population at large separations
than the ground-based proper motions and photometric parallaxes that
have been used in all studies so far.

Large, well-characterized samples of wide binaries have many
applications \citep{chaname2007}. In particular, the distribution of
wide binary semi-major axes can be used to constrain the properties
of molecular clouds and other massive structures in the disk, and
possible compact objects (MACHOs) in the dark halo
\citep{bahcalletal1985,yooetal2004}. If the distribution of binaries
can be measured at separations as large as a few parsecs we expect
to see ``tidal tails'' of the kind that have been detected around
globular clusters \citep{oden2001,belo06,gril06}; the evolution of
these structures offers a prototype for the evolution of the
phase-space structures in the solar neighborhood caused by the
disruption of stellar clusters \citep{dehnenbinney1998}.

Almost all theoretical studies of the expected distribution of wide
binaries have made two related approximations that compromise their
validity at the largest semi-major axes:

\begin{itemize}

\item The stars are assumed to disappear instantaneously as soon as their
  orbits become unbound. This is unrealistic because the disruption rate is
  dominated by weak, distant encounters, so most escaping stars
  have very small relative velocity and only drift slowly apart.

\item The Galactic tidal field is ignored. The tidal field becomes stronger
  than the gravitational attraction between the stars in the binary when the
  separation is roughly the Jacobi or tidal radius, which equals
  $1.7\pc=3.5\times10^5\au$ for solar-mass stars in the solar neighborhood (eq.\
  \ref{eq:rjacdef}). Thus the tidal field is already significant at the
  separations ($\sim 10^5\au$) probed by current measurements of the
  wide binary distribution, and dominates the dynamics at larger separations.

\end{itemize}

Including these two effects is necessary if we are to understand the
expected distribution of binary stars---bound and unbound---at
semi-major axes of $10^4\au$ and larger. To achieve this
understanding is the primary goal of this paper. We restrict
ourselves to the evolution of disk binaries under the influence of
passing stars, although it is straightforward to extend our methods
to include either halo binaries or other perturbers such as
molecular clouds or massive black holes.

The structure of this paper is as follows. In \S\ref{sec:basic}, we
describe the basic equations of motion for binary stars in the
Galactic tidal field and how we calculate the perturbations from
other stars that drive the orbital evolution.  We also review the
standard analytic treatment of the evolution of bound binaries using
a diffusion equation. Then in \S\ref{simulation} we describe the
results from our simulations.  Finally, \S\ref{sec:disc} contains a
discussion and conclusions, and Appendix \ref{diffEBS} derives an
analytic model that approximately describes the diffusion of unbound
binary stars.

\section{Basic equations in the numerical simulation}

\label{sec:basic} In this section, we describe the details of our
numerical simulation. First, we give the equations of motion of the
binary star in Hill's approximation. Then, we describe how we
include the effects of kicks from other stars. Finally, we describe
the diffusion approximation, which should be valid for binaries with
small semi-major axes.

\subsection{Evolution without kicks}

\label{sec:nokick}

We use Hill's approximation
\citep[e.g.,][]{heggie2001,binneytremaine2008} to describe the
motion of the binary star in the Galaxy. Hill's approximation is
valid because the mass of the binary is much less than the mass of
the Galaxy (by a factor $\sim 10^{11}$). Let the masses of the two
stars in the binary be $M_1$ and $M_2$. We assume that the potential
of the Galaxy is symmetric about the plane $Z=0$, where $(X,Y,Z)$ or
$(R,\phi,z)$ is an inertial Cartesian or cylindrical coordinate
system with origin at the center of the Galaxy. We introduce a
second coordinate system $(x,y,z)$ with origin in the $Z=0$ plane at
distance $R_g$ from the Galactic center. The $x$-$y$ and $X$-$Y$
planes coincide but the origin of the $(x,y,z)$ coordinate system
co-rotates with the Galaxy. The $x$-axis points radially outward,
the $y$-axis points in the direction of Galactic rotation, and the
$z$-axis is perpendicular to the Galactic plane. The $x$, $y$ and
$z$ axes form a right-hand coordinate system. With these
conventions, the positive $z$-axis points toward the South Galactic
Pole. The angular speed of the Galaxy at radius $R_g$, which equals
the angular speed of the $(x,y,z)$ frame, is $\mathbf{\Omega}_g=
\Omega_g\mathbf{e}_z$ in the $z$ or ``vertical'' direction. The
angular speed $\Omega_g$ is related to the potential of the Galaxy
$\Phi_0(R,z)$ by
\begin{equation}
\Omega_g^2=\frac{\Phi_0^{'}(R_g,0)}{R_g}, \label{Gphi}
\end{equation}
where $\Phi_0^{'}(R_g,0)\equiv\partial\Phi_0/\partial R|_{(R_g,0)}$.
As we want to study binary stars in the solar neighborhood, we can
just choose $R_g$ to be the distance of the Sun from the Galactic
center, $R_g=8\kpc$.

In the co-rotating frame with origin at $R_g$, the position of star
$i$, $i=1,2$, is labeled by $\mathbf{r}_i=(x_i,y_i,z_i)$. Then in
the co-rotating frame with origin at the center of the Galaxy, the
star's position is $\mathbf{R}_g+\mathbf{r}_i$, where
$\mathbf{R}_g=(R_g,0,0)$ and the equation of motion for either star
in the binary system is
\begin{equation}
\frac{d^2(\mathbf{r}_i+\mathbf{R}_g)}{dt^2}=-\bfnabla_i\Phi-
2\mathbf{\Omega}_g\times\frac{d(\mathbf{r}_i+\mathbf{R}_g)}{dt}
-\mathbf{\Omega}_g\times\left[\mathbf{\Omega}_g\times
(\mathbf{r}_i+\mathbf{R}_g)\right]\ , \label{origin1}
\end{equation}
where $\bfnabla_i$ is the gradient with respect to $\mathbf{r}_i$.
The potential $\Phi$ includes the contribution from the Galaxy
$\Phi_0$ as well as the potential of the binary stars $\Phi_b$. For
$\Phi_0$, we use the distant-tide approximation, which means
$\Phi_0(R_g+x,y,z)$ at the position of a star is expanded with
respect to $\Phi_g(R_g,0,z)$. Then we have
\begin{equation}
(\bfnabla \Phi_0)_\alpha
=\Phi_{0,\alpha}+\sum_{\beta=x,y,z}\Phi_{0,\alpha\beta}\
\beta+O(r^2),\quad \alpha=(x,y) .
\end{equation}
Here $\Phi_{0,\alpha}\equiv(\partial\Phi_0/
\partial \alpha)_{(R_g,0)}$ and $\Phi_{0,\alpha\beta}\equiv(\partial^2\Phi_0/
\partial \alpha\,\partial \beta)_{(R_g,0)}$. Equation (\ref{origin1}) is correct
for any value of $\mathbf{r}_i$; in particular, when
$\mathbf{r}_i=0$, it is correct for the center $\mathbf{R}_g$, and
we may subtract the equation for the center from (\ref{origin1}) to
obtain
\begin{align}
\frac{d^2 x}{dt^2}=&-\frac{\partial\Phi_b}{\partial x}
-\sum_{\beta=x,y,z}\Phi_{0,x\beta}\,\beta-
\left[2\mathbf{\Omega}_g\times\frac{d\mathbf{r}}{dt}\right]_x
-\left[\mathbf{\Omega}_g\times\left(\mathbf{\Omega}_g\times
\mathbf{r}\right)\right]_x, \nonumber \\
\frac{d^2 y}{dt^2}=&-\frac{\partial\Phi_b}{\partial y}
-\sum_{\beta=x,y,z}\Phi_{0,y\beta}\,\beta-
\left[2\mathbf{\Omega}_g\times\frac{d\mathbf{r}}{dt}\right]_y
-\left[\mathbf{\Omega}_g\times\left(\mathbf{\Omega}_g\times
\mathbf{r}\right)\right]_y, \nonumber \\
\frac{d^2z}{dt^2}=&-\frac{\partial\Phi_b}{\partial z}
-\frac{\partial\Phi_0}{\partial z}-
\left[2\mathbf{\Omega}_g\times\frac{d\mathbf{r}}{dt}\right]_{z}
-\left[\mathbf{\Omega}_g\times\left(\mathbf{\Omega}_g\times
\mathbf{r}\right)\right]_{z}.
\end{align}

We have
\begin{equation}
\Phi_{0,xx}=\Big(\frac{\partial^2\Phi_0}{\partial
R^2}\Big)_{(R_g,0)},\quad
\Phi_{0,yy}=\Big(\frac{1}{R}\frac{\partial\Phi_0}{\partial
R}\Big)_{(R_g,0)}, \quad \Phi_{0,xy}=0.
\end{equation}
The potential $\Phi_{b}$ for star $i=1,2$ is just the potential from
the other star in the binary. Then we have
\begin{equation}
\Phi_{b1}=-\frac{GM_2}{\sqrt{(x_1-x_2)^2+(y_1-y_2)^2+(z_1-z_2)^2}},
\label{binarypotential}
\end{equation}
where $(x_i,y_i,z_i)$ denotes the position of star $i$, and the
formula for $\Phi_{b2}$ is obtained by interchanging 1 and 2 in all
subscripts. Then the equations of motion for either star are
\begin{align}
\ddot{x_i}=&2\Omega_g
\dot{y_i}+\Big[\Omega_g^2-\Phi_0^{''}(R_g,0)\Big]x_i-
\frac{\partial \Phi_{bi}}{\partial x_i}\ ,\nonumber\\
\ddot{y_i}=&-2\Omega_g
\dot{x_i}+\Big[\Omega_g^2-\frac{\Phi_0^{'}(R_g,0)}{R_g}\Big]y_i-
\frac{\partial \Phi_{bi}}{\partial y_i}\ ,\nonumber\\
\ddot{z_i}=&-\frac{\partial \Phi_0}{\partial
z_i}-\frac{\partial\Phi_{bi}}{\partial z_i}\ . \label{motion1}
\end{align}
As the angular speed $\Omega_g$ is related to the potential $\Phi_0$
via equation (\ref{Gphi}), we have
\begin{equation}
\Phi_0^{''}(R_g,0)=\Omega_g^2+2R_g\Omega_g\frac{d\Omega}{dR}\bigg|_{R_g}.
\end{equation}
As usual, the Oort constant $A(R)$ is defined as
\begin{equation}
A(R)=-\frac{1}{2}R\frac{d\Omega}{dR}\ .
\end{equation}
We label $A_g=A(R_g)$. Then equation (\ref{motion1}) can be
simplified to
\begin{align}
\ddot{x_i}-2\Omega_g \dot{y_i}-4\Omega_gA_gx_i=&-\frac{\partial
  \Phi_{bi}}{\partial x_i}\ ,\nonumber\\
\ddot{y_i}+2\Omega_g \dot{x_i}=&-\frac{\partial \Phi_{bi}}{\partial
y_i}\
,\nonumber\\
\ddot{z_i}+\frac{\partial \Phi_0}{\partial
  z_i}=&-\frac{\partial\Phi_{bi}}{\partial z_i}\ .
\label{motion2}
\end{align}

{}From equation (\ref{binarypotential}), we have the following
relations
\begin{equation}
M_1\frac{\partial\Phi_{b1}}{\partial
x_1}=-M_2\frac{\partial\Phi_{b2}}{\partial x_2},\quad
M_1\frac{\partial\Phi_{b1}}{\partial
y_1}=-M_2\frac{\partial\Phi_{b2}}{\partial y_2},\quad
M_1\frac{\partial\Phi_{b1}}{\partial
z_1}=-M_2\frac{\partial\Phi_{b2}}{\partial z_2}.
\end{equation}
The center of mass of the binary system $\mathbf{r}_{\rm{cm}}$ is
defined to be
\begin{equation}
\mathbf{r}_{\rm{cm}}=\frac{M_1\mathbf{r}_1+M_2\mathbf{r}_2}{M_1+M_2}.
\end{equation}
The relative coordinates of the two stars are
$\mathbf{r}=\mathbf{r}_1-\mathbf{r}_2.$ By adding equations
(\ref{motion2}) multiplied by appropriate coefficients for star
$i=1$ and $i=2$, we get the equation for the motion of the center of
mass\footnote{We assume the separation of the two stars along the
$z$ direction is much smaller than the thickness of the Galaxy.}
\begin{align}
\ddot{x}_{\rm{cm}}-2\Omega_g \dot{y}_{\rm{cm}}-4\Omega_gA_gx_{\rm{cm}}=&0\ ,\nonumber\\
\ddot{y}_{\rm{cm}}+2\Omega_g \dot{x}_{\rm{cm}}=&0\ ,\nonumber\\
\ddot{z}_{\rm{cm}}+\frac{\partial \Phi_0}{\partial z_{\rm{cm}}}=&0\
. \label{motioncm}
\end{align}
Due to the symmetry of the Galactic potential,
$(\partial\Phi_0/\partial z)_{z=0}=0$, so for stars not very far
from the mid-plane of the Galaxy, we approximately have $\partial
\Phi_0/\partial z=(\partial^2\Phi_0/\partial z^2)_{z=0}\,z$ and we
define
\begin{equation}
\nu_g^2=\frac{\partial^2\Phi_0}{\partial z^2}\bigg{|}_{(R_g,0)},
\end{equation}
where $\nu_g$ is the frequency for small oscillations in $z$. The
general solution to the above equations of motion for the center of
mass is just epicycle motion,
\begin{align}
x_{\rm{cm}}(t)=&x_{g,\rm cm}+X\cos(\kappa_gt+\alpha)\ ,\nonumber\\
y_{\rm{cm}}(t)=&y_{g,\rm cm}(t)-Y\sin(\kappa_gt+\alpha),\quad
y_{g,\rm cm}(t)=y_{g,0}-2A_gx_{g,\rm cm}t,
\quad Y=\frac{2\Omega_g}{\kappa_g}X,\nonumber\\
z_{\rm{cm}}(t)=&Z\cos(\nu_gt+\alpha_z). \label{solutioncm}
\end{align}
Here $x_{g,\rm cm}$, $X$, $y_{g,0}$, $Z$, $\alpha$, $\alpha_z$ are
arbitrary constants, and $\kappa_g$ is the epicycle frequency
defined by
\begin{equation}
\kappa_g^2=4\Omega_g(\Omega_g-A_g)\ .
\end{equation}
The variables $x_{g,\rm cm}(t)$ and $y_{g,\rm cm}(t)$ give the
position of the guiding center---the center of the epicyclic
motion---for the center of mass. Subtract equations (\ref{motion2})
with $i=2$ from $i=1$ and we get the equations for the relative
motion of the two stars
\begin{align}
\ddot{x}-2\Omega_g \dot{y}-4\Omega_gA_gx=&-\frac{G(M_1+M_2)x}{(x^2+y^2+z^2)^{3/2}}\ ,\nonumber\\
\ddot{y}+2\Omega_g \dot{x}=&-\frac{G(M_1+M_2)y}{(x^2+y^2+z^2)^{3/2}}\ ,\nonumber\\
\ddot{z}+\nu_g^2 z=&-\frac{G(M_1+M_2)z}{(x^2+y^2+z^2)^{3/2}}\ .
\label{solutionrela}
\end{align}
In this set of equations, the terms involving $\dot{y}$ or $\dot{x}$
arise from the Coriolis force, as we are working in a rotating
frame; the terms involving $A_g$ or $\nu_g$ represent the effect of
the Galactic tide, and the terms on the right side represent the
gravitational force between the members of the binary system.

Equations (\ref{solutionrela}) show that the relative motion is the
same as that of a test particle around an object with the mass
$M_1+M_2$ in the Galactic tidal field. A special solution to the
above equations is the stationary solution
($\ddot{x}=\dot{x}=\ddot{y}= \dot{y}=\ddot{z}=\dot{z}=0$)
\begin{equation}
y=z=0,\ x=\pm r_J, \quad \hbox{where}\quad
r_J\equiv\left[\frac{G(M_1+M_2)}{4\Omega_gA_g}\right]^{1/3}\
\label{stationarysolution}
\end{equation}
is the Jacobi or tidal radius of the binary system. The stationary
points are actually the Lagrange points in the three-body system
composed of binary star and the Galaxy. As will be seen from our
simulations below, the Jacobi radius sets the characteristic scale
for the distribution of binary stars at large radii.

Equations (\ref{solutionrela}) admit one integral of motion, the
Jacobi constant
\begin{align}
E_J\equiv&\frac{1}{2}(\dot{x}^2+\dot{y}^2+\dot{z}^2-4\Omega_gA_gx^2+\nu_g^2z^2)
-\frac{G(M_1+M_2)}{\sqrt{x^2+y^2+z^2}}\nonumber\\
=&\frac{1}{2}(\dot{x}^2+\dot{y}^2+\dot{z}^2)+\Phi_{\rm{eff}}(x,y,z),
\end{align}
where $\Phi_{\rm{eff}}(x,y,z)\equiv \nu_g^2z^2/2-2\Omega_gA_gx^2
-G(M_1+M_2)/\sqrt{x^2+y^2+z^2}$ is the effective potential. The
Jacobi constant for the stationary solution
(\ref{stationarysolution}) is called the critical Jacobi constant
$E_{c}$, and is given by
\begin{equation}
E_{c}=\Phi_{\rm{eff}}(\pm r_J,0,0)=-2\Omega_g A_g
r_J^2-\frac{G(M_1+M_2)}{r_J}=-\frac{3}{2^{1/3}}(\Omega_gA_g)^{1/3}[G(M_1+M_2)]^{2/3}.
\label{eq:ecdef}
\end{equation}
As $\dot{x}^2,\dot{y}^2,\dot{z}^2\ge0$, the motion is constrained to
the region in which $\Phi_{\rm eff}(x,y,z) \le E_J$, and the
boundary of this region is the zero-velocity surface for a given
Jacobi constant, defined implicitly by
\begin{equation}
\Phi_{\rm{eff}}(x,y,z)=E_J.
\end{equation}

\begin{figure}
\includegraphics[bb=0 125 638 666,width=\hsize]{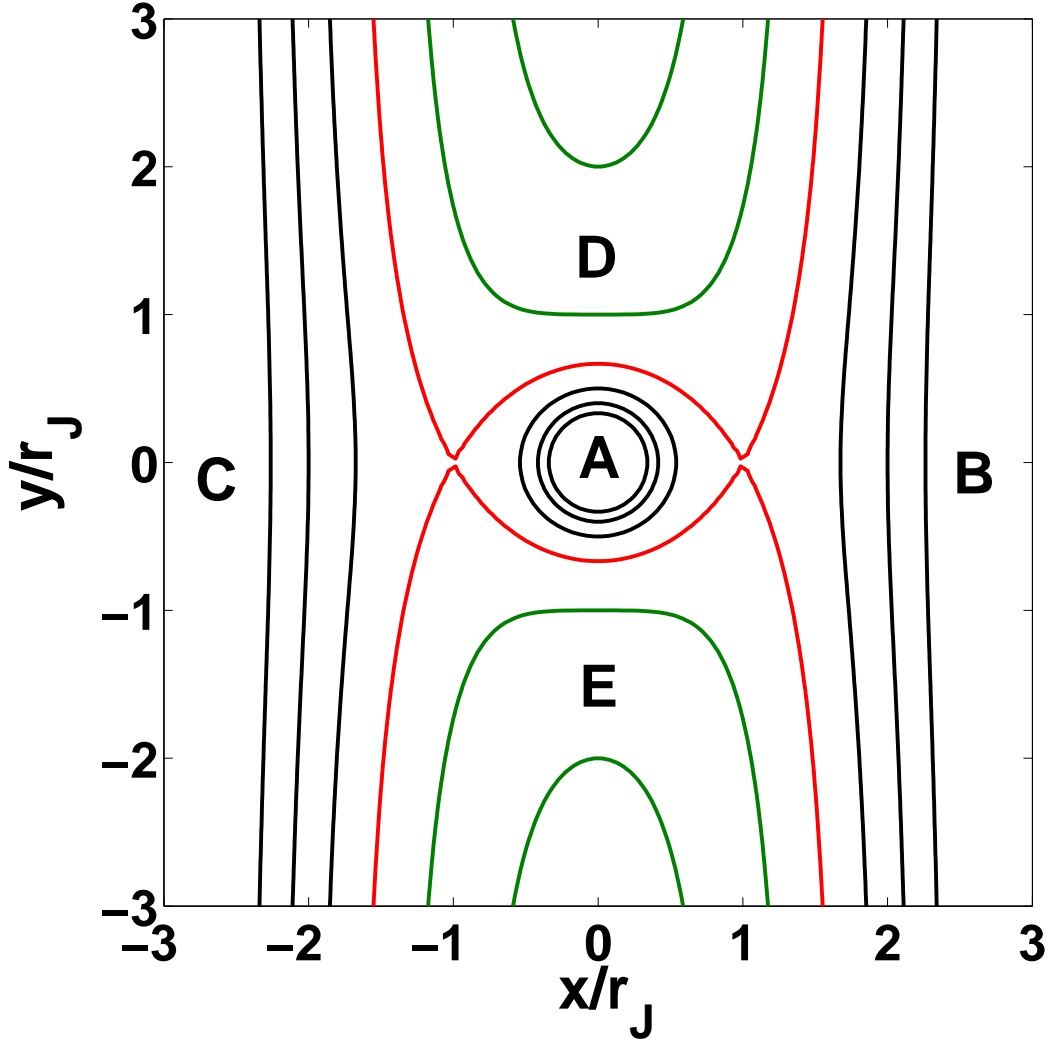}\\
\caption{Zero-velocity contours in the plane parallel to the
Galactic disk, as given by equation (\ref{contourequa}).  The red
line is the critical contour on which the effective potential
$\Phi_{\rm{eff}}=E_c$. In regions A, B and C, $\Phi_{\rm{eff}}<E_c$
and in regions D and E, $\Phi_{\rm{eff}}>E_c$. Binary stars with
large separations can either have $\Phi_{\rm{eff}}<E_c$ or
$\Phi_{\rm{eff}}>E_c$, depending on their positions on this plot. We
define bound binaries as those in region A with $E_J<E_c$ and call
all others escaped binaries.} \label{contour}
\end{figure}

We choose the time unit to be $1/\Omega_g$ and the length unit to be
$r_J$. Then we can define the following dimensionless variables (see
eq.\ \ref{eq:vjacdef} for numerical values of the scaling factors)
\begin{equation}
\tilde{\mathbf{r}}=\frac{\mathbf{r}}{r_J},\quad
\tilde{\mathbf{r}}^{\prime}=\frac{\dot{\mathbf{r}}}{\Omega_g
r_J},\quad
\tilde{\mathbf{r}}^{\prime\prime}=\frac{\ddot{\mathbf{r}}}{\Omega_g^2r_J}.
\end{equation}
Then equations (\ref{solutionrela}) can be simplified to the
following dimensionless form
\begin{align}
\tilde{x}^{\prime\prime}-2\tilde{y}^{\prime}-\frac{4A_g}{\Omega_g}\tilde{x}=&
-\frac{4A_g}{\Omega_g}\frac{\tilde{x}}{(\tilde{x}^2+\tilde{y}^2+\tilde{z}^2)^{3/2}}\ ,\nonumber\\
\tilde{y}^{\prime\prime}+2\tilde{x}^{\prime}=&-\frac{4A_g}{\Omega_g}\frac{\tilde{y}}{(\tilde{x}^2+\tilde{y}^2+\tilde{z}^2)^{3/2}}\ ,\nonumber\\
\tilde{z}^{\prime\prime}+\frac{\nu_g^2}{\Omega_g^2}
\tilde{z}=&-\frac{4A_g}{\Omega_g}\frac{\tilde{z}}{(\tilde{x}^2+\tilde{y}^2+\tilde{z}^2)^{3/2}}\
. \label{dimrela}
\end{align}
Note that the dimensionless equations do not depend on the specific
values of the masses $M_1$ and $M_2$. Thus the result applies to
binaries of any masses. The dimensionless form of the zero-velocity
surface projected to the $x$-$y$ plane is
\begin{equation}
\tilde{x}^2+\frac{2}{\sqrt{\tilde{x}^2+\tilde{y}^2}}=-\frac{E_J}{2r_J^2A_g\Omega_g}.
\label{contourequa}
\end{equation}
The zero-velocity contours are shown in Figure \ref{contour}.
Binaries with $E_J<E_c$ in region A have bounded motion in that they
can never escape from A; we call these bound binaries. All others
are called escaped binaries.

Once we know the velocity of the center of mass
$\dot{\mathbf{r}}_{\rm{cm}}$ and relative velocity
$\dot{\mathbf{r}}$, we can calculate the velocity of each star in
the rotating frame from the relations
\begin{equation}
\dot{\mathbf{r}}_{1}=\dot{\mathbf{r}}_{\rm{cm}}+\frac{M_2}{M_1+M_2}\dot{\mathbf{r}},
\quad
\dot{\mathbf{r}}_{2}=\dot{\mathbf{r}}_{\rm{cm}}-\frac{M_1}{M_1+M_2}\dot{\mathbf{r}}.
\label{eq:vdef}
\end{equation}

\subsection{Kicks from other stars}
\label{kick}

In order to study the evolution of the binary systems, we must
include the effect of encounters with passing stars and other
perturbers (e.g., molecular clouds).  In this paper we only discuss
the effects of encounters with stars, but we return briefly to the
effects of molecular clouds in the discussion of \S\ref{sec:disc}.

As the velocity dispersion of the perturbers ($\sim30\kms$) is much
larger than the velocity difference of the two stars in the binary
system ($\lesssim1\kms$), we can use the impulse approximation,
i.e., the encounter with the perturber provides an impulsive kick
that changes only the velocity, not the position, of the subject
star.  For computational efficiency, we do not follow individual
encounters but instead consider the total effect of the encounters
on the binary system after some time interval $\Delta t_p$, which is
generally large enough to include many encounters (see
\S\ref{valuesofparameters} for further discussion of this
approximation).  Let the change of velocity of the subject star
after this time interval be $\Delta \mathbf{v}$. According to the
central limit theorem, the effect of a large number of kicks will be
the same as that of a Gaussian distribution with the same mean $
\boldsymbol{\mu}=\langle\Delta \mathbf{v}\rangle$ and covariance
matrix $C_{\alpha\beta}=\langle\Delta v_{\alpha}\Delta
v_{\beta}\rangle$, where the subscripts $\alpha,\ \beta$ refer to
the $x$, $y$, $z$ directions. The values of $\mu_{\alpha}$ and
$C_{\alpha\beta}$ after the time interval $\Delta t_p$ can be
computed from the diffusion coefficients
\begin{equation}
\mu_{\alpha}=D[\Delta v_{\alpha}]\Delta t_p,\quad
C_{\alpha\beta}=D[\Delta v_{\alpha}\Delta v_{\beta}]\Delta t_p\ .
\end{equation}

We assume that the number density of perturbers with mass in the
range $M_p\rightarrow M_p+dM_p$ is $n(M_p)dM_p$, and that the
velocity distribution of the perturbers relative to the center of
mass of the binary is isotropic and Maxwellian,
\begin{equation}
dn=f(v_p)dM_pd\mathbf{v}_p=\frac{n(M_p)}{(2\pi\sigma^2)^{3/2}}
e^{-v_p^2/(2\sigma^2)}dM_pd\mathbf{v}_p, \label{fvp}
\end{equation}
where $\sigma$ is the relative velocity dispersion. Expressions for
the diffusion coefficients are given in equations (7.89) and (7.92)
of \cite{binneytremaine2008}. The actual relative velocity
distribution is more complicated, both because the distribution of
stellar velocities in the solar neighborhood is triaxial and because
the center of mass of the subject binary star has its own epicyclic
motion, but we do not believe that these complications will alter
our results significantly. If the velocity of subject star $i$
relative to the center of mass of the binary is $\mathbf{v}_i$, then
\begin{align}
D[\Delta v_{i,\alpha}]=&\frac{v_{i,\alpha}}{v_i}D[\Delta v_{||}]_i,\ \nonumber\\
D[\Delta v_{i,\alpha}\Delta
v_{i,\beta}]=&\frac{v_{i,\alpha}v_{i,\beta}}{v_i^2}\big\{ D[(\Delta
v_{||})^2]_i-\frac{1}{2}D[(\Delta v_{\perp})^2]_i\big\}
+\frac{1}{2}\delta_{\alpha,\beta}D[(\Delta v_{\perp})^2]_i,
\label{kickcoefficient}
\end{align}
where $D[\Delta v_{||}]$ is the mean change of velocity per unit
time along the velocity vector direction $\hat{\mathbf{v}}$, while
$D[(\Delta v_{\perp})^2]$ and $D[(\Delta v_{\parallel})^2]$ are the
mean-square changes per unit time in the velocity perpendicular and
parallel to $\hat{\mathbf{v}}$.

In the limit $|\mathbf{v}_i|\ll\sigma$, we have
\begin{align}
D[\Delta v_{||}]_i=&\frac{4\sqrt{2\pi} G^2(M_i\rho_{1}+\rho_{2})\ln\wedge}{3\sigma^3}v_i,\ \nonumber\\
D[(\Delta v_{||})^2]_i=&\frac{8\sqrt{2\pi} G^2 \rho_{2}\ln
\wedge}{3\sigma}\ ,\nonumber\\
D[(\Delta v_{\perp})^2]_i=&\frac{16\sqrt{2\pi} G^2 \rho_{2} \ln
\wedge}{3\sigma}, \label{diffcoef}
\end{align}
where
\begin{equation}
\rho_{k}=\int n(M_p)M_p^k\,dM_p. \label{eq:rhodef}
\end{equation}
Here $\wedge$ is defined to be
\begin{equation}
\wedge=\frac{b_{\rm{max}}v_{\rm{typ}}^2}{G(M_i+\tilde{M}_p)},
\end{equation}
where $b_{\rm{max}}$ is the maximum impact parameter considered,
$v_{\rm{typ}}$ is the typical relative velocity and $\tilde{M}_p$ is
the typical perturber mass. Since these parameters enter only
logarithmically, we can just assume $\tilde{M}_p=\msun$ and
$v_{\rm{typ}}\simeq\sigma$ for simplicity, and the maximum impact
parameter $b_{\rm{max}}$ can be chosen to be the half of the
separation of the two stars when we apply the kick\footnote{We have
checked that even though the separations of binary stars have a wide
range, different choices of $b_{\rm{max}}$ will not change the
results significantly.}.


By the central limit theorem the distribution function for $\Delta
\mathbf{v}$ is
\begin{equation}
f(\Delta
\mathbf{v})=\frac{1}{(2\pi)^{3/2}|\mathbf{C}|^{1/2}}\exp\left[-\frac{1}{2}(
\Delta \mathbf{v}-\boldsymbol{\mu})^{\top}\cdot \mathbf{M}\cdot
(\Delta \mathbf{v}-\boldsymbol{\mu})\right]. \label{distribution}
\end{equation}
Here, $\Delta \mathbf{v}$ and $\boldsymbol{\mu}$ are $3\times1$
matrices and $\mathbf{M}=\mathbf{C}^{-1}$ is a $3\times3$ matrix.
Then the evolution of the binary system is followed numerically by
repeating the following steps: (i) follow the orbital evolution for
a time interval $\Delta t_p$ using the equations of motion
(\ref{dimrela}); (ii) for each of the two stars, draw a random kick
velocity $\Delta \mathbf{v}$ from the distribution
(\ref{distribution}) and add this kick velocity to the velocity of
the star.

\subsection{The diffusion approximation for small semi-major axes}
\label{diffusion}

When the semi-major axis $a$ of the binary system is small enough
($a\ll r_J$), the effect of the Galactic tide is small compared with
the mutual gravitational force of the two stars. Then the binary
evolves as an isolated two-body system subject to kicks from other
stars.  Moreover the energy kicks from passing stars are small
compared to the binding energy of the binary, so the evolution can
be treated using the diffusion approximation.  This problem has been
studied by previous researchers
\citep[e.g.,][]{king1977,rk1982,Weinbergetal1987}, so we just give
the equations here. We will use this diffusion approximation both to
speed up calculations of the binary evolution at small semi-major
axis and to provide insight into the numerical results.

The energy of the binary system $E$ is related to the semi-major
axis $a$ by $E=-G(M_1+M_2)/(2a)$. We define $\tilde{n}(E,t)dE$ to be
the number of binary systems with energy in the range $[E,\ E+dE]$
at time $t$. The diffusion equation reads \citep[][eq.\
B1]{Weinbergetal1987}
\begin{equation}
\frac{\partial \tilde{n}(E,t)}{\partial
t}=\epsilon\left\{-\frac{\partial \tilde{n}(E,t)}{\partial E}
-\frac{2}{3}\frac{\partial^2}{\partial
E^2}\left[E\tilde{n}(E,t)\right]\right\}\ , \label{diffCBS}
\end{equation}
where
\begin{equation}
\epsilon=8\pi
G^2\rho_{2}\left\langle\frac{1}{V_{\rm{rel}}}\right\rangle\ln
\wedge\ .
\end{equation}
Here $\left\langle1/V_{\rm{rel}}\right\rangle$ is the average
inverse relative velocity between the binary and the perturbers,
which is $\sqrt{2/\pi}\sigma^{-1}$ under the assumption that the
relative velocity distribution is given by (\ref{fvp}). As
$\epsilon$ depends on the binary energy very weakly (through
$\wedge$), we take it to be independent of $E$. We define two
dimensionless variables $\tau$ and $h$ by
\begin{equation}
\tau=\frac{2\epsilon t}{3|E_1|},\quad h=-\frac{E}{|E_1|}\ ,
\end{equation}
where $E_1$ is a scaling parameter. Then
$\tilde{n}(E,t)dE=\tilde{n}(h,\tau)dh$. With the boundary condition
$\tilde{n}(E_1,t)=0$ and the initial condition
$\tilde{n}(E,0)=\delta(E-E_0)$, the solution to the diffusion
equation (\ref{diffCBS}) is \citep[][eq.\ B13]{Weinbergetal1987}
\begin{equation}
\tilde{n}(h,\tau)=512\pi
h_0^{5/2}\int_0^{\infty}dk\frac{e^{-k\tau}k^5F_{k}(h_0)F_k(h)}{9+12k+16k^2},
\end{equation}
where $h_0=-E_0/|E_1|$ and
\begin{equation}
F_k(h)=\left(2\sqrt{kh}\right)^{-5/2}\left[J_{5/2}\left(2\sqrt{kh}\right)
J_{-5/2}\left(2\sqrt{k}\right)
-J_{5/2}\left(2\sqrt{k}\right)J_{-5/2}\left(2\sqrt{kh}\right)\right],
\end{equation}
with $J_{\nu}(z)$ the Bessel function of order $\nu$. Then the
probability that the energy of the binary system is larger than
$E_1$ for the first time in the interval ($t,t+dt$) is $p(t)dt$,
where
\begin{equation}
p(t)=-\frac{\partial}{\partial t}\int_1^{\infty}dh\
\tilde{n}(h,\tau)=
\frac{2^{15/2}h_0^{5/2}\epsilon}{3|E_{1}|}\int_0^{\infty}dk\frac{e^{-k\tau}k^{15/4}F_k(h_0)}{9+12k+16k^2}\
.\label{escapetime}
\end{equation}
To use these results to accelerate our calculation, we choose two
semi-major axes $a_0$ and $a_1$, which are small enough ($a_0<a_1\ll
r_J$) that the influence of the Galactic tide is negligible. We set
$E_{0}$ and $E_{1}$ to be the corresponding binary energies.
Typically $a_0=0.5 a_1$ in our simulation. If the semi-major axis of
the binary system in our Monte Carlo simulation random walks to a
value smaller than $a_0$ at time $t_0$, then we draw a random time
$t$ from the distribution function (\ref{escapetime}), which is a
fair sample of the time the binary system needs to go from $a_0$ to
$a_1$. If $t_0+t$ is smaller than the total time of our simulation
(10 Gyr), then we just give the binary system semi-major axis $a_1$
and other randomly chosen orbital elements as described at the start
of the following section, and continue to evolve the binary
numerically from time $t_0+t$. If its semi-major axis becomes less
than $a_0$ a second time, we just repeat the above calculation. If
$t_0+t$ is larger than the total time of the simulation, then we
conclude that at the end of our simulation, the semi-major axis of
the binary system is still smaller than $a_1$. Then the probability
for the binary system to have dimensionless energy $[h,h+dh]$ at
time $t$ is
\begin{equation}
p_2(h,\tau)dh=\frac{\tilde{n}(h,\tau)dh}{\int_1^{+\infty}\tilde{n}(h,\tau)dh}.
\label{eq:probdiff}
\end{equation}
Here the time $t$ (and thus $\tau$) is fixed by the condition that
$t_0+t$ is the time at the end of the simulation. We draw a random
number from this distribution and use this to determine the
semi-major axis of the binary system at the end of the simulation.
We include this binary system in the final statistical result after
we assign random values to the other orbital elements of the orbit
as described at the start of the following section.

\section{Numerical simulation of the binary systems}
\label{simulation}

We simulate the evolution of binary systems for up to 10 Gyr under
the influence of the Galactic tide and kicks from passing stars. To
determine the initial relative position and velocity of the two
stars in the binary system, we first choose a semi-major axis $a_i$
as described below. We then choose the inclination angle $\theta$
between the plane of the orbit and the Galactic plane randomly so
that $\cos\theta$ is uniformly distributed between $-1$ and $1$,
which corresponds to a spherical distribution. We choose the
eccentricity $e$ of the initial orbit so that $e^2$ is distributed
uniformly random between $0$ and $1$, which corresponds to an
ergodic distribution on the energy surface. The angle between the
projected major axis of the orbit on the $x$-$y$ plane and the
$x$-axis is uniformly distributed between $0$ and $2\pi$. The
initial phase or mean anomaly of the orbit is also uniformly
distributed between $0$ and $2\pi$.

We carry out six simulations of $N=50,000$ binary stars each. In the
first four simulations the systems are ``formed'' at initial times
$t_0$ that are uniformly distributed between 0 and 10 Gyr and
followed to $t=10$ Gyr, to represent the current state of a
population with uniform star-formation rate. The initial semi-major
axes in units of the Jacobi radius are
$a_i/r_J=0.01,0.05,0.10,0.20$. In the final two simulations the
logarithms of the initial semi-major axes are uniformly distributed
(\"Opik's law, eq.\ \ref{eq:opikinit}) between $\log_{10}(0.001r_J)$
and $\log_{10}(0.5r_J)$; in the first of these the binary formation
times $t_0$ are uniformly distributed between 0 and 10 Gyr, while in
the second the binaries are all formed at $t_0=0$. We label these
``\"Opik 1'' and ``\"Opik 2''. In these cases a small fraction
($<7\%$) of the initial binaries have already escaped in that
$E_J>E_c$.

Initially, the center of mass is on a circular orbit, which means
$\mathbf{r}_{\rm{cm}}=\mathbf{\dot{r}}_{\rm{cm}}=0$.

We solve the equations of motion (\ref{dimrela}) numerically over
the time interval $\Delta t_p$ using an adaptive fourth-order
Runge-Kutta method and Kustaanheimo-Stiefel regularization
\citep{StiefelScheifele1971}. The evolution of the center of mass
over this interval is given by the solution (\ref{solutioncm}). At
the end of this interval, we know the velocities and positions of
the two stars. Then we generate random velocity kicks $\Delta
\mathbf{v}_i$ for each star from the distribution function
(\ref{distribution}).  We add $\Delta \mathbf{v}_i$ to each star
while keeping the positions unchanged. With the new velocities and
positions as initial conditions, we let the binary system evolve for
another time interval $\Delta t_p$. If the semi-major axis becomes
smaller than a specified value $a_0$, we switch to the diffusion
approximation as described in \S\ref{diffusion} until either (i) we
reach 10 Gyr and stop, or (ii) the semi-major axis exceeds
$a_1>a_0$, at which point we return to a numerical
simulation.\footnote{For the simulations with
$a_i/r_J=0.05,0.1,0.2$, we chose $a_0/r_J=0.04,0.05,0.1$ and
$a_1=2a_0$. In the simulation with $a_i/r_J=0.01$, we initially
followed the evolution of all stars using the diffusion
approximation and switched to the Monte Carlo simulation when the
semi-major axis exceeded $a_1/r_J=0.08$. In the \"Opik 1 and \"Opik
2 simulations our procedure depended on the initial semi-major axis:
for $a_i/r_J>0.2$ we did not use the diffusion approximation at all;
for  $0.08\le a_i/r_J<0.2$, we used $a_1=2a_0=a_i$; for
$a_i/r_J<0.08$, we initially followed the evolution using the
diffusion approximation and switched to the simulation at
$a_1/r_J=0.08$.} If the initial semi-major axis $a_i$ is smaller
than $a_0$ we start with the diffusion approximation. We follow each
binary system in this way to the time 10 Gyr. If we are using the
diffusion approximation at 10 Gyr, we draw a random semi-major axis
from the probability distribution (\ref{eq:probdiff}) and assign the
other orbital elements at random as described at the beginning of
this section.

\subsection{Values of the parameters in the numerical simulation}
\label{valuesofparameters}

In this subsection, we give the values of the parameters we chose in
the simulation. As we focus on binaries in the solar neighborhood,
the angular speed $\Omega_g$, vertical frequency $\nu_g$, Oort
constant $A_g$, and epicycle frequency $\kappa_g$ are chosen to be
the values in the solar neighborhood, taken from Table 1.2 of
\cite{binneytremaine2008}
\begin{align}
\Omega_g=&\,236\kms/ 8\kpc=9.56\times10^{-16}\ \mbox{s}^{-1}\ ,\nonumber\\
\nu_g=&\,2.3\times10^{-15}\,\mbox{s}^{-1}\ ,\ \nonumber\\
A_g=&\,14.8\kms\kpc^{-1}=4.796\times10^{-16}\ \mbox{s}^{-1}\ ,\nonumber\\
\kappa_g=&\,37\kms\kpc^{-1}=1.2\times10^{-15}\ \mbox{s}^{-1}\ .
\end{align}
With these units, the Jacobi radius
\begin{equation}
r_J=1.70\pc\left(\frac{M_1+M_2}{2\msun}\right)^{1/3}.
\label{eq:rjacdef}
\end{equation}
The corresponding velocity and acceleration are
\begin{equation}
\Omega_gr_J=0.050\kms\left(\frac{M_1+M_2}{2\msun}\right)^{1/3},
\quad \Omega_g^2r_J=4.8\times 10^{-17}\,\mbox{km
  s}^{-2}\left(\frac{M_1+M_2}{2\msun}\right)^{1/3}.
\label{eq:vjacdef}
\end{equation}

The typical one-dimensional velocity dispersion in the solar
neighborhood is $29\kms$ \citep{dehnenbinney1998} and we choose the
relative velocity dispersion to be $\surd{2}$ times this, so
$\sigma=40\kms$.


The mass function $n(M_p)$ of the stars in the solar neighborhood is
given in equation (1) of \cite{kroupaetal1993}:
\begin{equation}
n(M_p)=n_0 \left\{ \begin{array}
{r@{\quad,\quad}l}0\quad\quad\quad & \mbox{if}\ \ M_p<M_{p,l},\\
\left(M_p/M_{p,0}\right)^{-\alpha_1} & \mbox{if}\ M_{p,l}\le
M_p<M_{p,0},
\\(M_p/M_{p,0})^{-\alpha_2} & \mbox{if}\ M_{p,0}\le M_p<M_{p,r},
\\ (M_{p,r}/M_{p,0})^{-\alpha_2}(M_p/M_{p,r})^{-\alpha_3} & \mbox{if}\ M_{p,r}\le M_p.
\end{array} \right.
\end{equation}
The parameters in this equation are
\begin{align}
\alpha_1=&1.3,\quad \alpha_2=2.2,\quad
\alpha_3=4.5, \nonumber\\
n_0=&0.087\pc^{-3}\msun^{-1}\ ,\nonumber\\
M_{p,l}=&0.07\msun,\quad M_{p,0}=0.5\msun,\quad M_{p,r}=1\msun.
\label{massfunction}
\end{align}
The moments of the mass function are then
\begin{align}
\rho_{0}=&\int n(M_p)dM_p=0.14\ \mbox{pc}^{-3},\nonumber\\
\rho_{1}=&\int n(M_p)M_pdM_p=0.045\msun\pc^{-3},\nonumber \\
\rho_{2}=&\int n(M_p)M_p^2dM_p=0.029\msun^2\pc^{-3}.
\label{massdensity}
\end{align}

Although the dimensionless equations of motion (\ref{dimrela}) and
the diffusion coefficients $D[(\Delta v_{||})^2]$ and $D[(\Delta
v_{\perp})^2]$ do not depend on the specific values of the binary
component masses $M_1$ and $M_2$, the diffusion coefficient
$D[\Delta v_{||}]$ (eq.\ \ref{diffcoef}) actually depends on these
masses. When we calculate this kick we choose $M_1=M_2=\msun$.

We now describe the choice of the interval $\Delta t_p$ between
kicks. If the two stars are bound, we can find the semi-major axis
$a$ from the energy equation
\begin{equation}
E=-\frac{G(M_1+M_2)}{2a}=\frac{1}{2}v^2-\frac{G(M_1+M_2)}{r}\ .
\end{equation}
Then the orbital frequency of the binary system $\Omega_b$ is
\begin{equation}
\Omega_b=\sqrt{\frac{G(M_1+M_2)}{a^3}},
\end{equation}
and the orbital period is
\begin{equation}
P_b=\frac{2\pi}{\Omega_b}=3\times10^6\yr\left(\frac{a}{0.1\pc}\right)^{3/2}.
\end{equation}
For a typical number density of stars in the solar neighborhood
$0.05\pc^{-3}$ and a typical relative velocity of the stars
$40\kms$, the collision time (time between encounters with impact
parameter less than $a$) between the binary system and the field
star is
\begin{equation}
t_{\rm{coll}}=1.25\times10^{7}\left(\frac{0.1\pc}{a}\right)^2\yr.
\label{eq:tcolldef}
\end{equation}
If the energy $E$ is positive, the time interval $\Delta t_p$ is
chosen to be
\begin{equation}
\Delta t_p=\frac{0.1}{\Omega_g}=3.3\times 10^6\yr.
\end{equation}
If the energy is negative and the collision time is longer than the
orbital period, the time interval $\Delta t_p$ is just chosen to be
the collision time.  If the collision time is shorter than the
period, $\Delta t_p$ is chosen to be
\begin{equation}
\Delta t_p=\frac{0.1}{\mbox{max}(\Omega_g,\Omega_b)}\ .
\label{eq:follw}
\end{equation}
We assumed in \S\ref{kick} that the interval $\Delta t_p$ was large
compared to the encounter time. This assumption is not correct for
bound binaries with semi-major axes $\lesssim 0.5\pc$. Nevertheless,
our results should accurately reproduce the evolution of the binary
so long as the evolution time is much longer than the encounter
time, since the central limit theorem implies that any distribution
of velocity kicks with the correct mean and covariance matrix should
lead to the same cumulative effects. We have checked this by varying
the value of $\Delta t_p$ by a factor of $3$, and found almost no
change in the final distribution of the binary systems.

We have also checked our simulation code by following 5000 binary
systems with initial semi-major axis $0.05r_J$, stopping each
simulation when the semi-major axis reaches $0.1r_J$. In this range
of semi-major axes the Galactic tidal force is at least $10^3$ times
smaller than the gravitational force between the binary components,
so the diffusion approximation given in \S\ref{diffusion} should be
quite accurate. We compared the cumulative distribution of stopping
times to the distribution predicted by the diffusion approximation
(eq.\ \ref{escapetime}) and the maximum difference was only 2\%.

\begin{figure}
\centering
\subfloat[]{\includegraphics[width=0.85\textwidth]{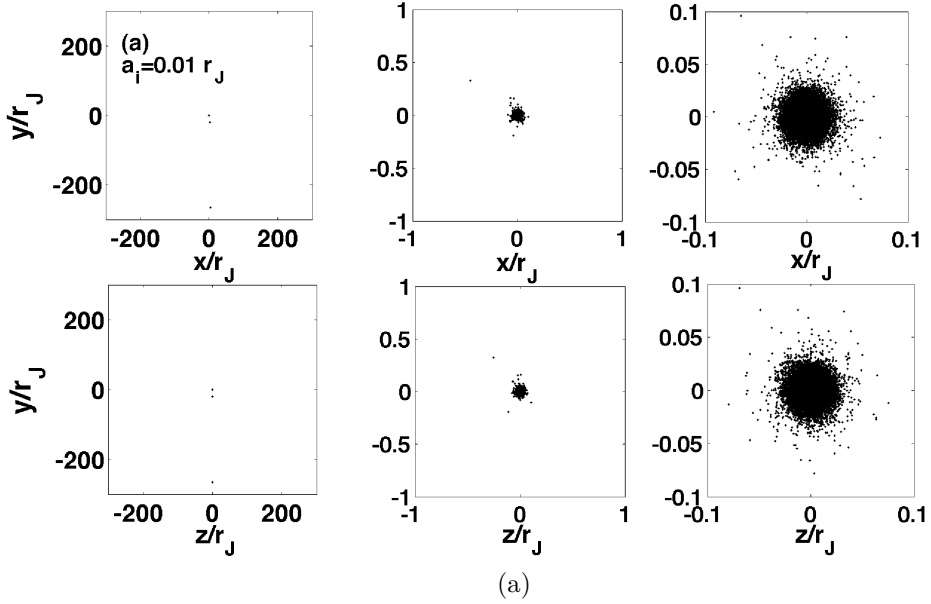}}\\
\vspace{5mm}
\subfloat[]{\includegraphics[width=0.85\textwidth]{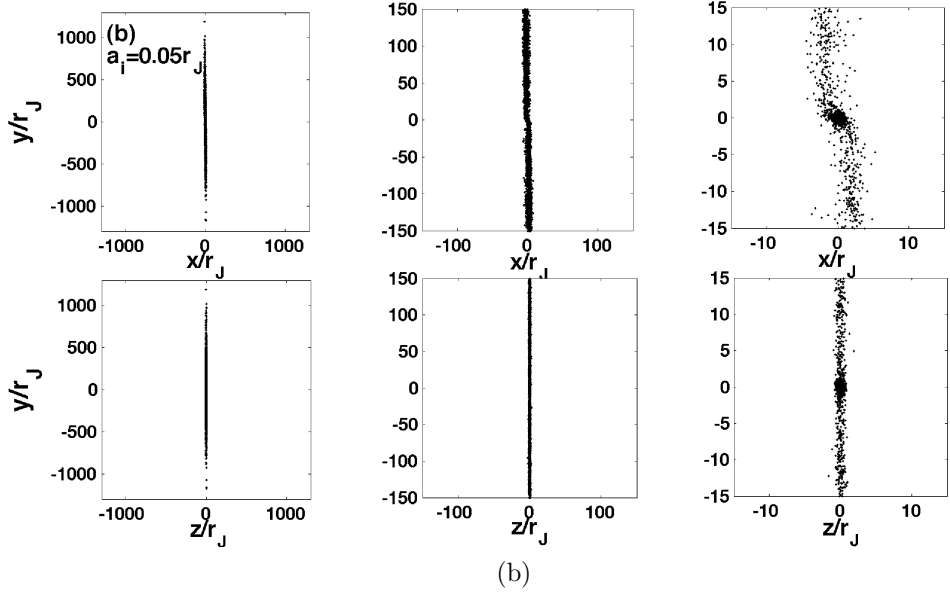}}\\
\caption{The relative position $\mathbf{r}_1-\mathbf{r}_2$ in 50,000
binary
  systems at time 10 Gyr with four different initial semi-major axes: (a)
  $0.01r_J=0.017\pc$, (b) $0.05r_J=0.085\pc$, (c) $0.1r_J=0.17\pc$, (d)
  $0.2r_J=0.34\pc$. In these simulations the binaries are formed at a uniform
  rate between $t=0$ and $t=10\Gyr$. We also show two simulations in which the
  initial semi-major axes are uniformly distributed in the log between
  $0.001r_J=0.0017\pc$ and $0.5r_J=0.85\pc$: (e) formation at a uniform rate
  between $t=0$ and $t=10\Gyr$; (f) all binaries formed at $t=0$.  The frames
  labeled $x$ and $y$ are distributions projected onto the $x$--$y$ plane
  (parallel to the Galactic plane) at various scales while the frames
  labeled $z$ and $y$ are distributions projected onto the $z$--$y$ plane. The
  larger the initial semi-major axis, the more stars are found in the tidal
  tails.} \label{space}
\end{figure}

\begin{figure}
\ContinuedFloat \centering
\subfloat[]{\includegraphics[width=0.85\textwidth]{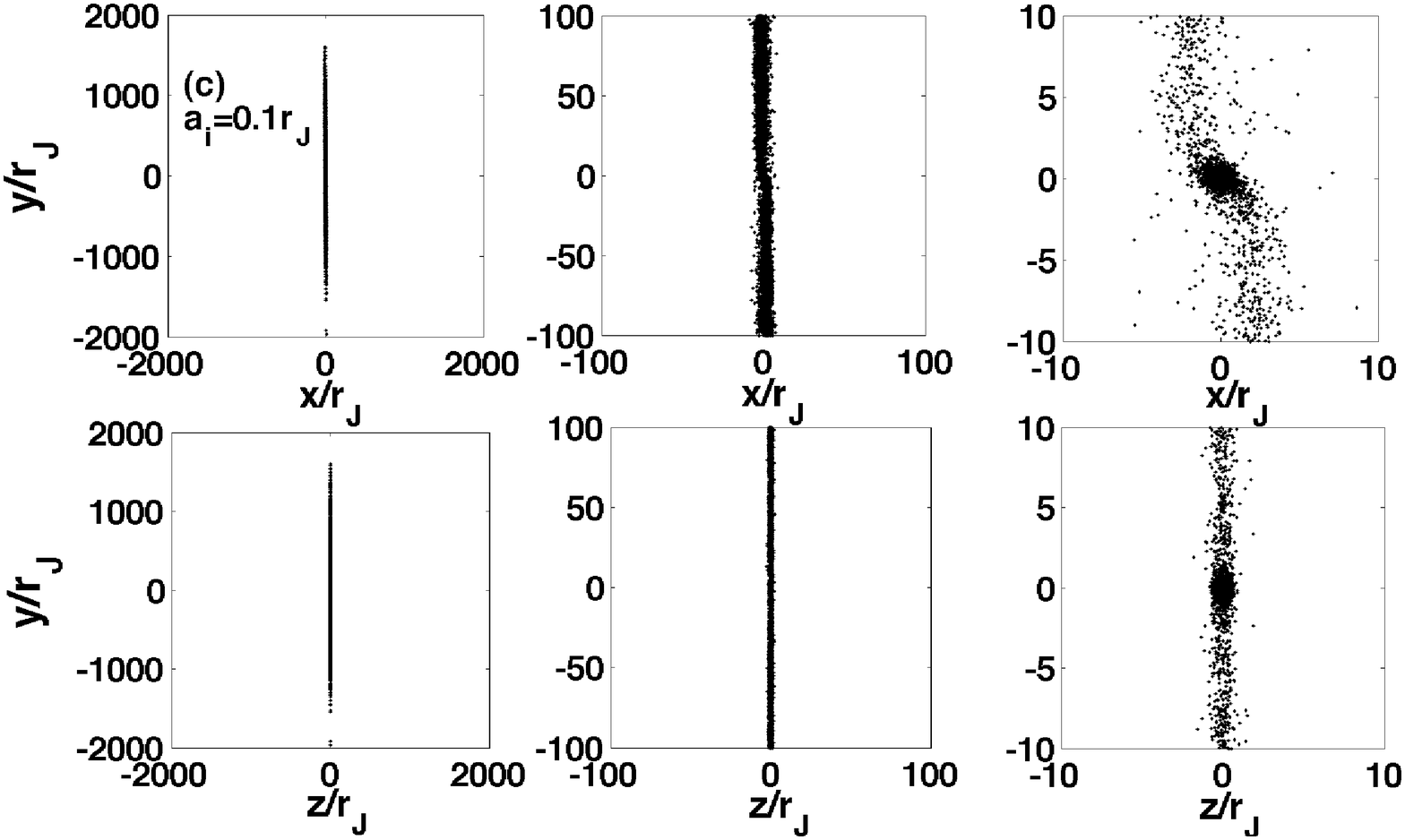}}\\
\vspace{20mm}
\subfloat[]{\includegraphics[width=0.85\textwidth]{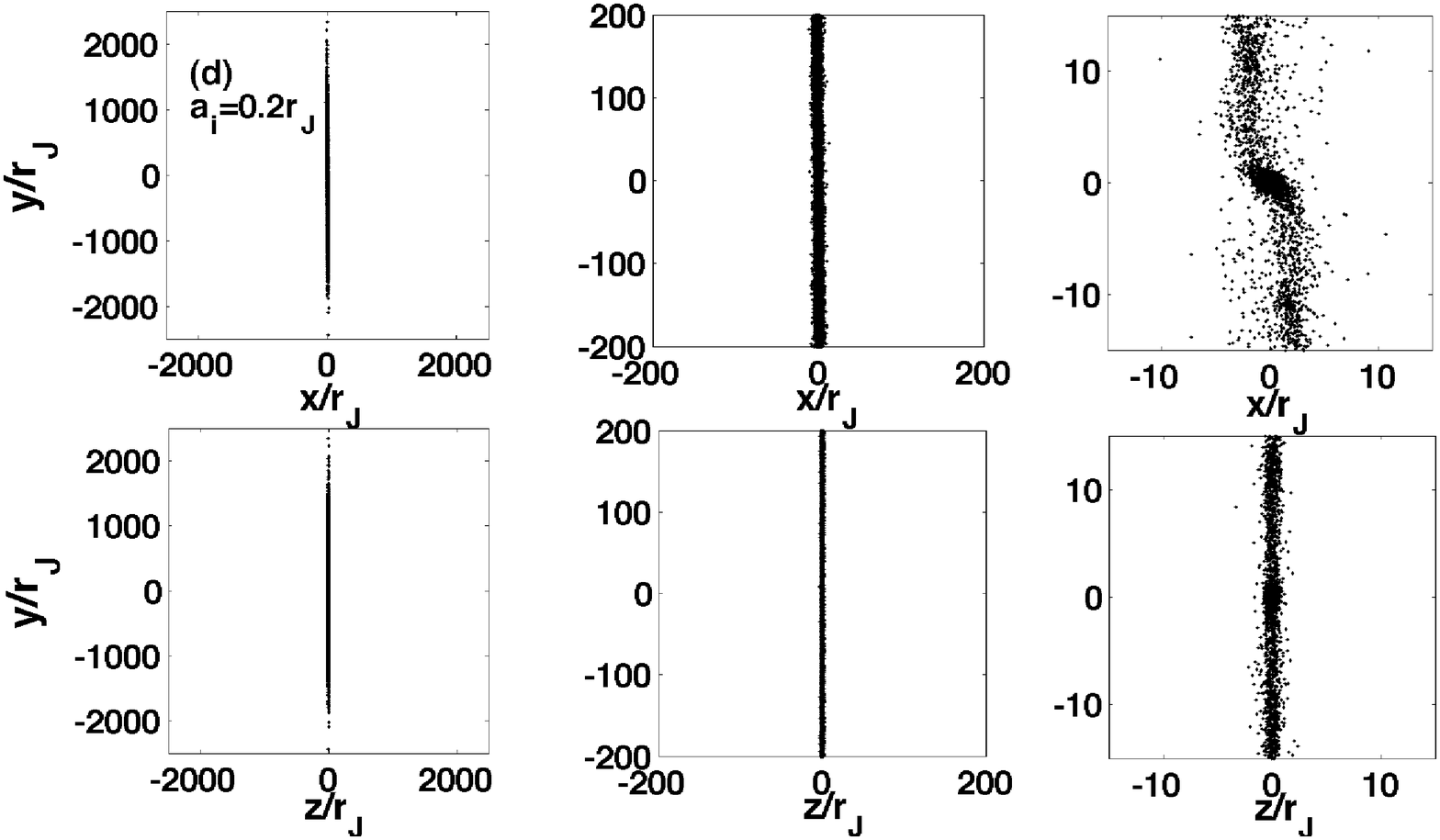}}\\
\caption{}
\end{figure}

\begin{figure}
\ContinuedFloat \centering
\subfloat[]{\includegraphics[width=0.85\textwidth]{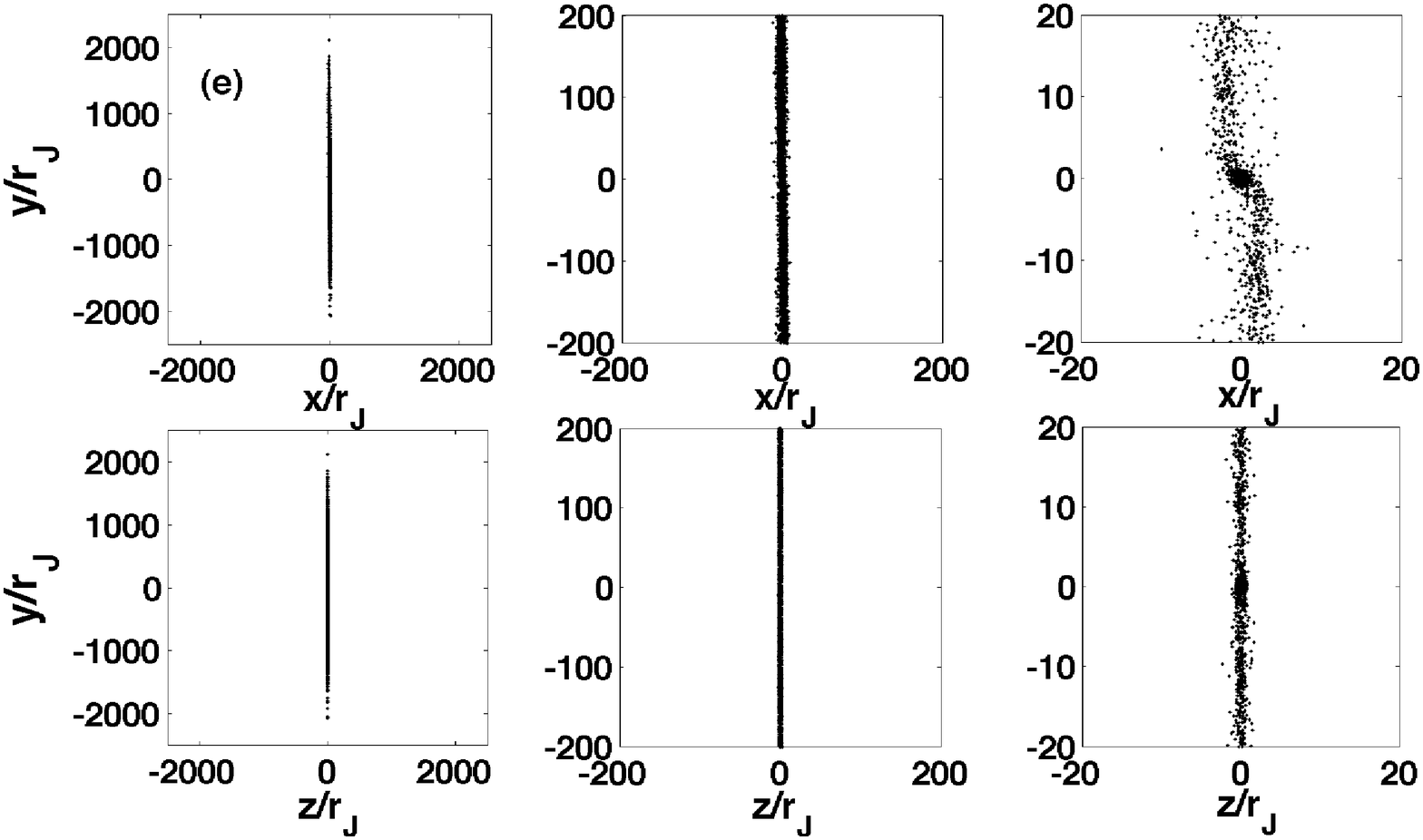}}\\
\vspace{20mm}
\subfloat[]{\includegraphics[width=0.85\textwidth]{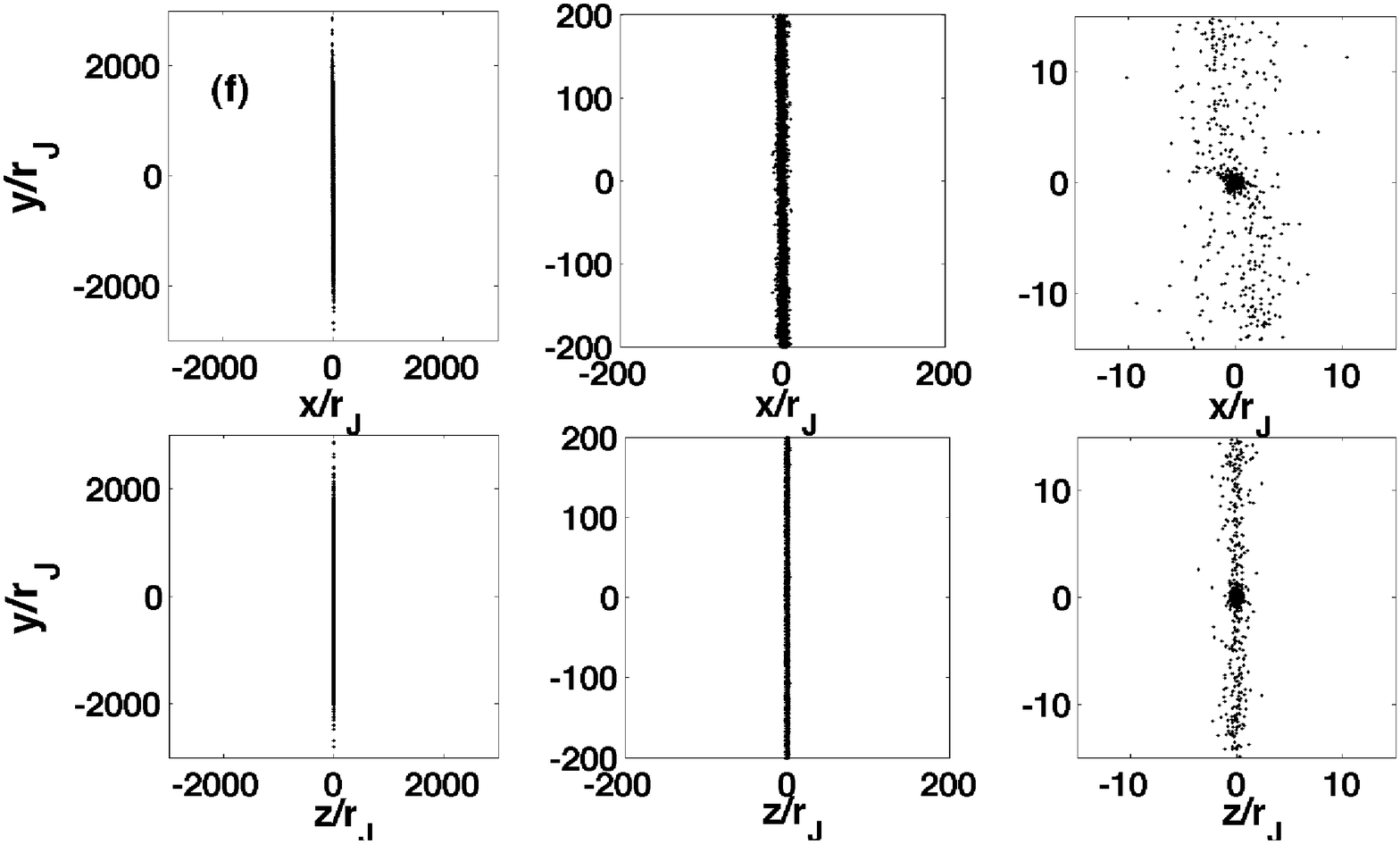}}\\
\caption{}
\end{figure}

\begin{figure}
\centering
\includegraphics[width=0.48\hsize]{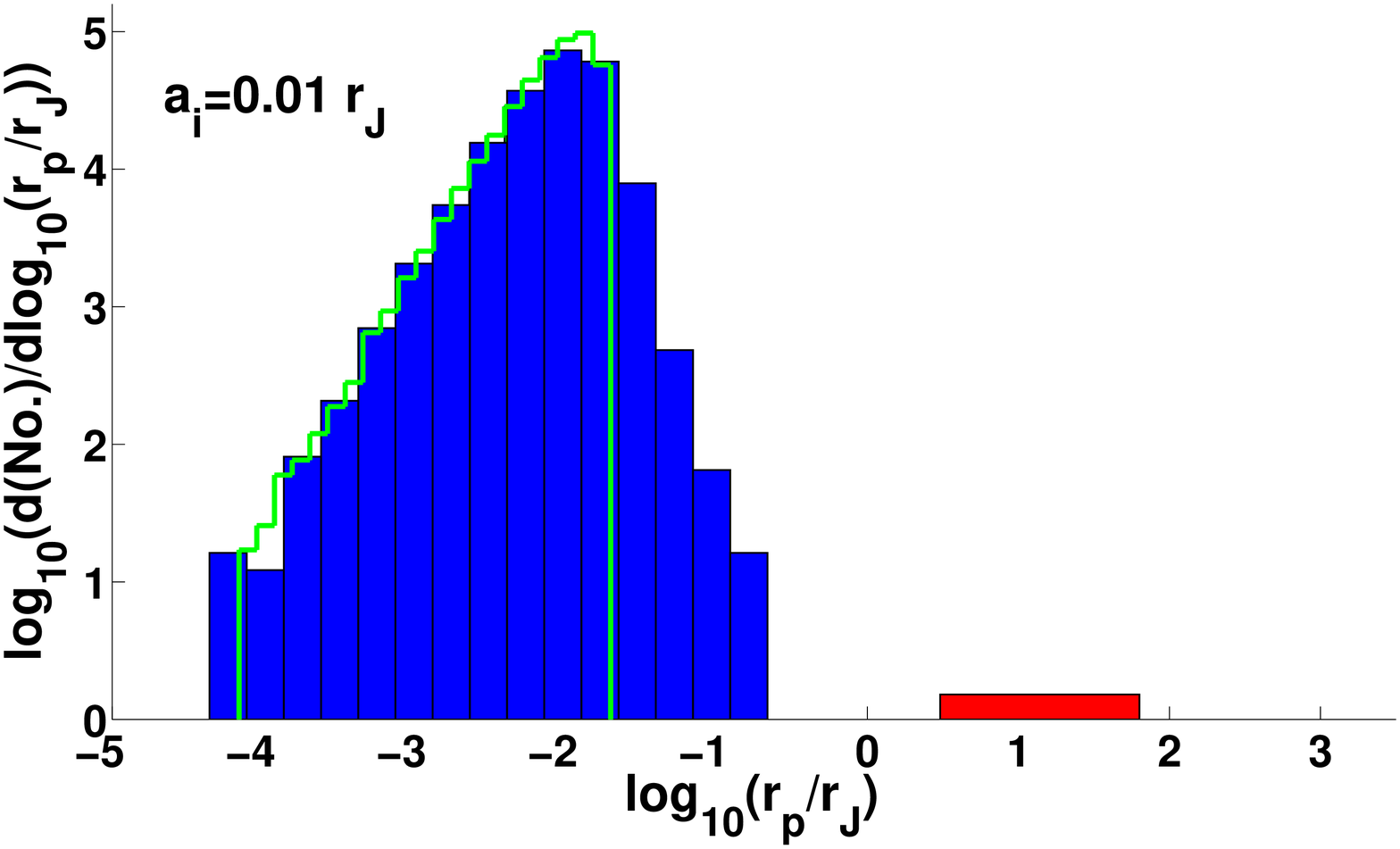}
\hfill
\includegraphics[width=0.48\hsize]{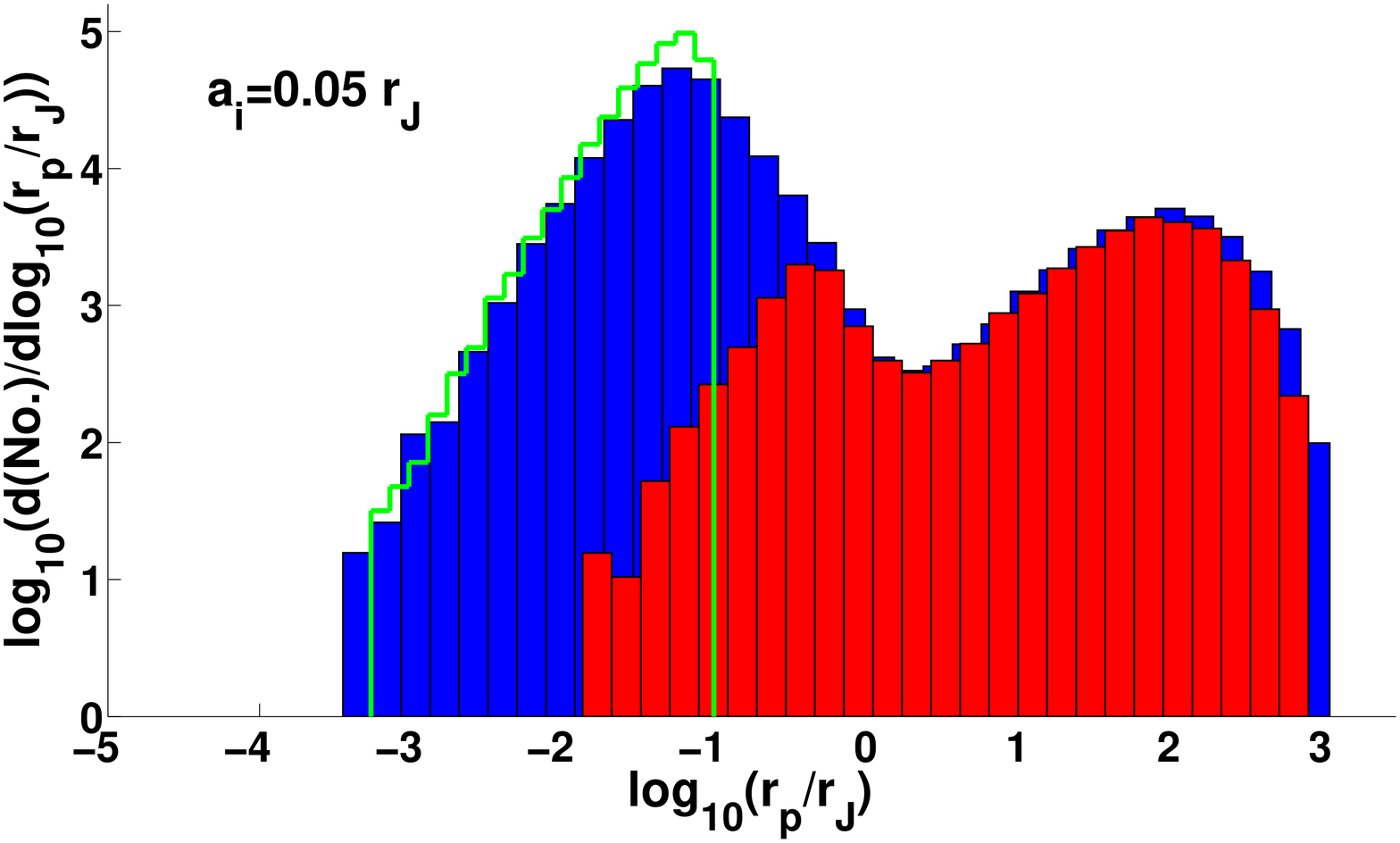} \\
\vspace{5mm}
\includegraphics[width=0.48\hsize]{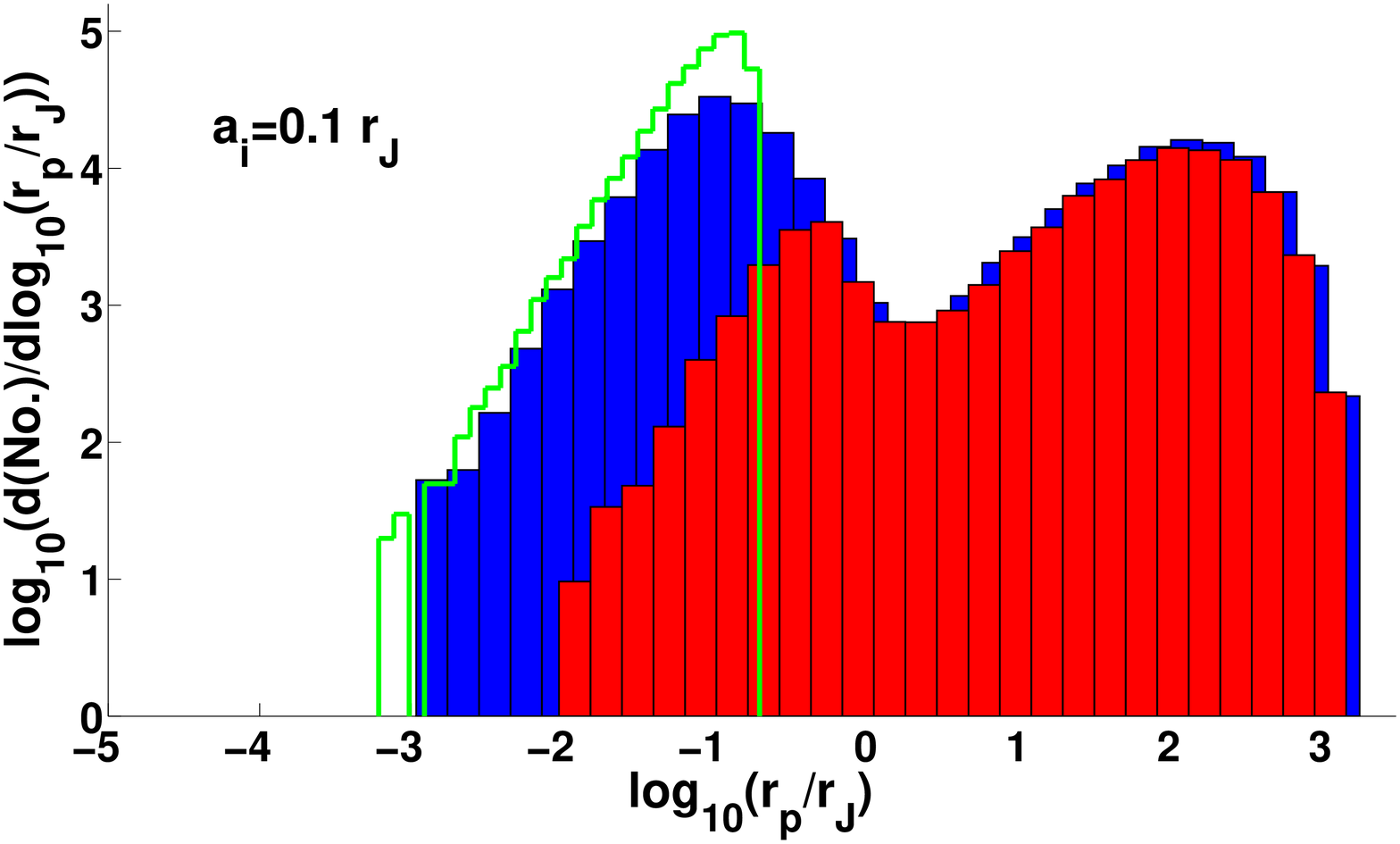}
\hfill
\includegraphics[width=0.48\hsize]{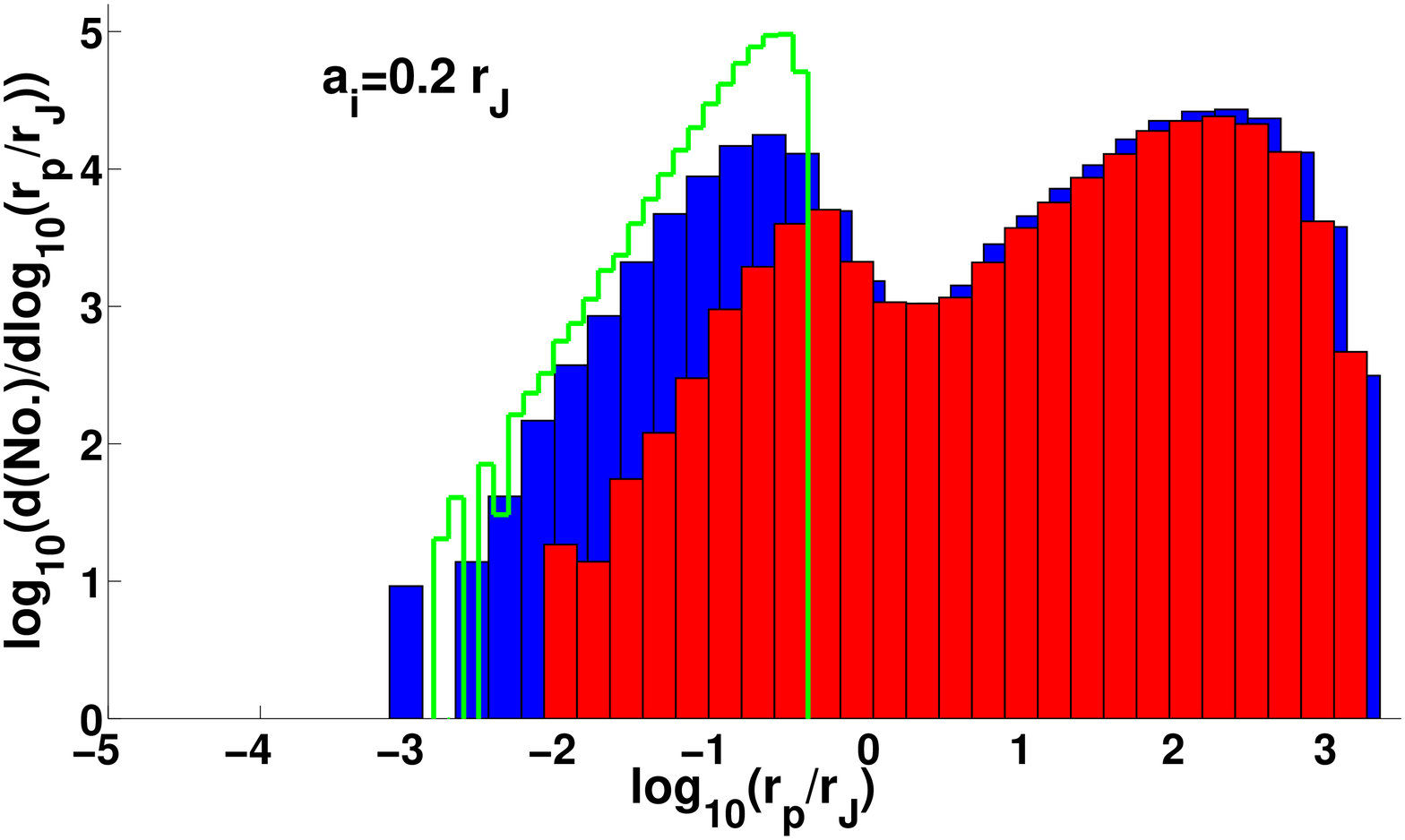} \\
\vspace{5mm}
\includegraphics[width=0.48\hsize]{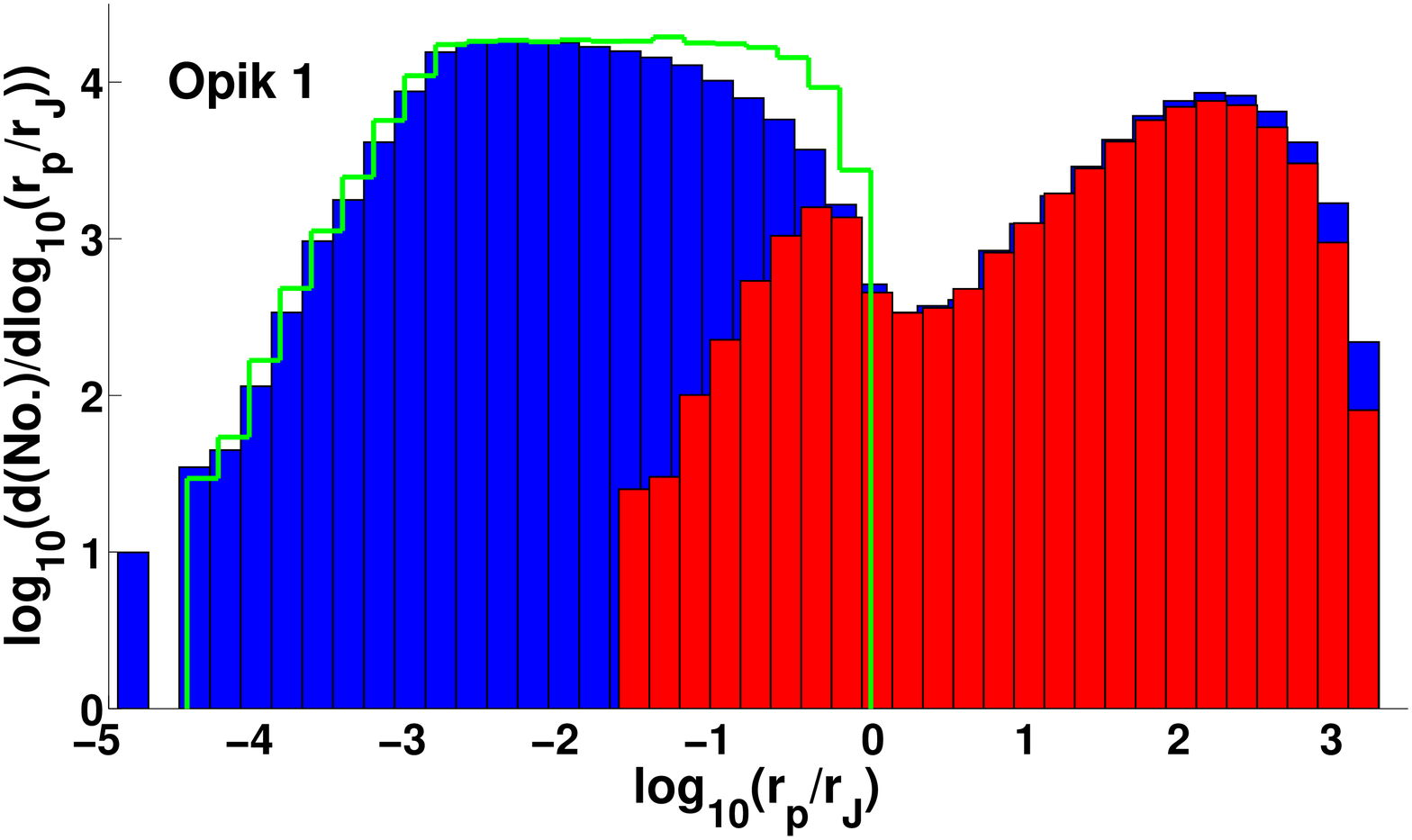}
\hfill
\includegraphics[width=0.48\hsize]{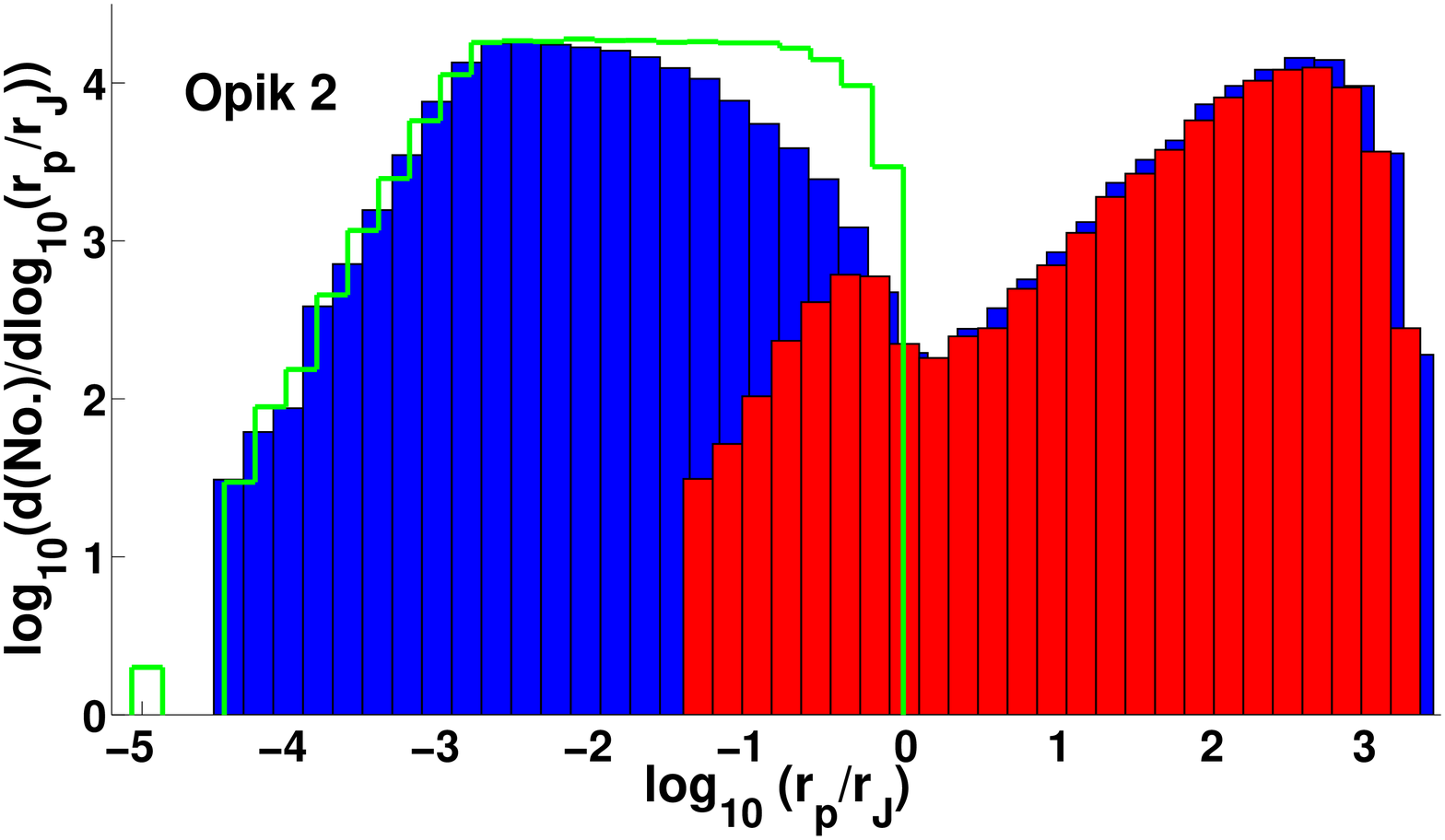} \\
\caption{Histograms of the projected separation $r_p$ of the binary
stars
  after 10 Gyr, for initial semi-major axes $a_i=0.01r_J$, $0.05r_J$,
  $0.1r_J$, $0.2r_J$ (top four panels) and two cases in which $\log_{10} a_i$ is
  uniformly distributed between $\log_{10}(0.001r_J)$ and $\log_{10}(0.5r_J)$: the
  initial time $t_0$ is uniformly distributed between $0$ and $10\Gyr$ (\"Opik
  1; bottom left panel), and the initial time $t_0$ is fixed to be $0$ (\"Opik
  2; bottom right panel). The projected separation is obtained by assuming
  that the line between the two stars has a random angle to the line of sight.
  The histograms for real three-dimensional separations $r$ are very similar
  to the histograms of projected separations shown here. The blue histograms
  show the total sample of $50,000$ binary stars while the red histograms show
  the stars with $E_J>E_c$ at time 10 Gyr. The initial distribution is shown
  in green. There is a minimum in the
  distribution near $5r_J$ in each case.} \label{phist}
\end{figure}

\begin{figure}
\includegraphics[width=\hsize]{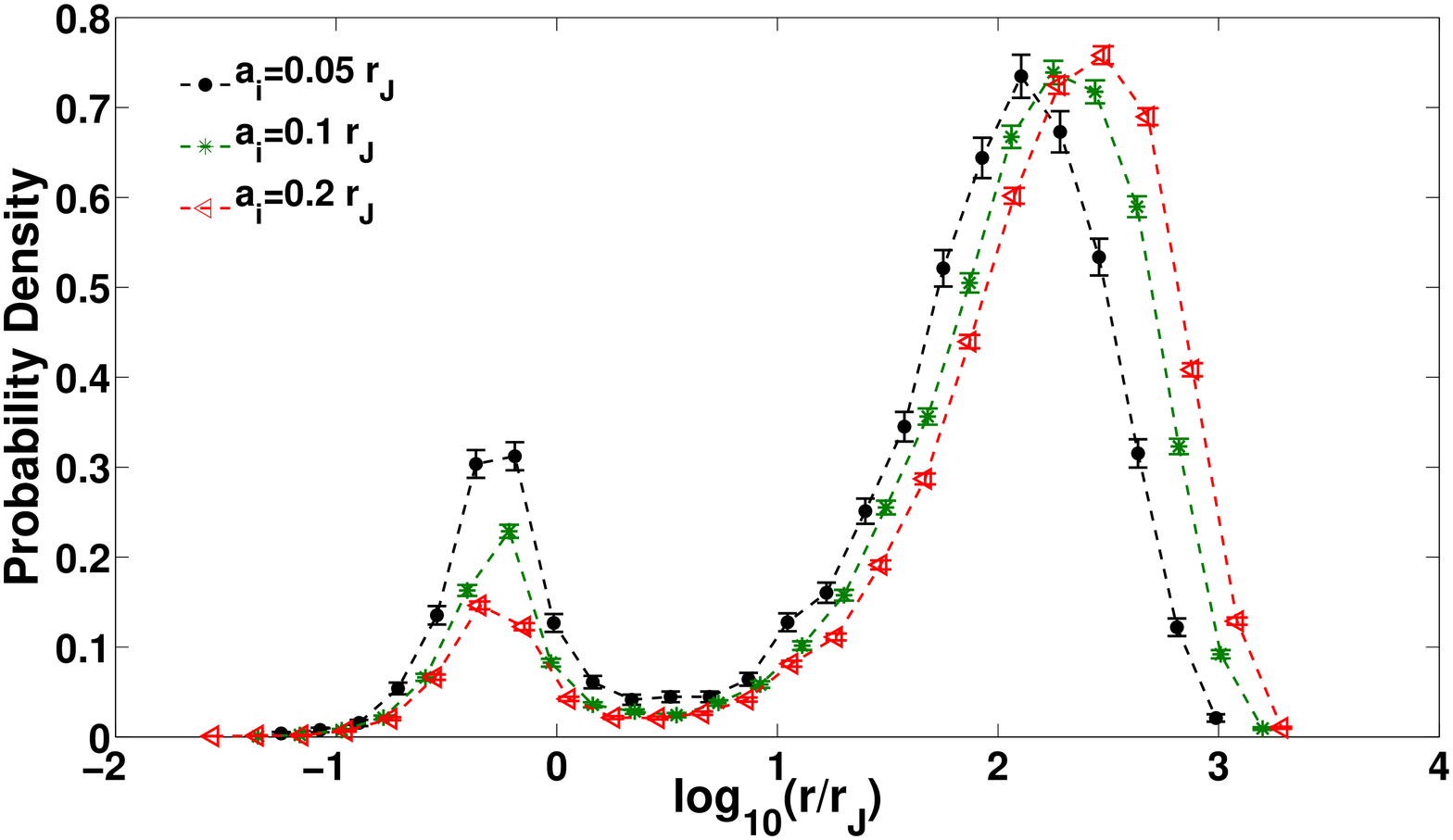}\\
\caption{Distribution of separations of escaped binary stars at 10
Gyr (the
  precise definition of ``escape'' is given in the caption of Figure
  \ref{contour}).  The figure shows the probability density of the escaped
  stars for the simulations with initial semi-major axis $a_i=0.05r_J$,
  $0.1r_J$, and $0.2r_J$, normalized so that the integral of the probability
  density over $\log_{10}(r/r_J)$ is unity. There are two peaks separated by a
  minimum around $5r_J$. The larger the initial semi-major axis, the larger is
  the amplitude and centroid of the exterior peak. The simulation for initial
  semi-major axis $a_i=0.01r_J$ is not shown because the number of escaped
  stars is too small.}\label{escaped}
\end{figure}

\begin{figure}
\includegraphics[width=0.49\hsize]{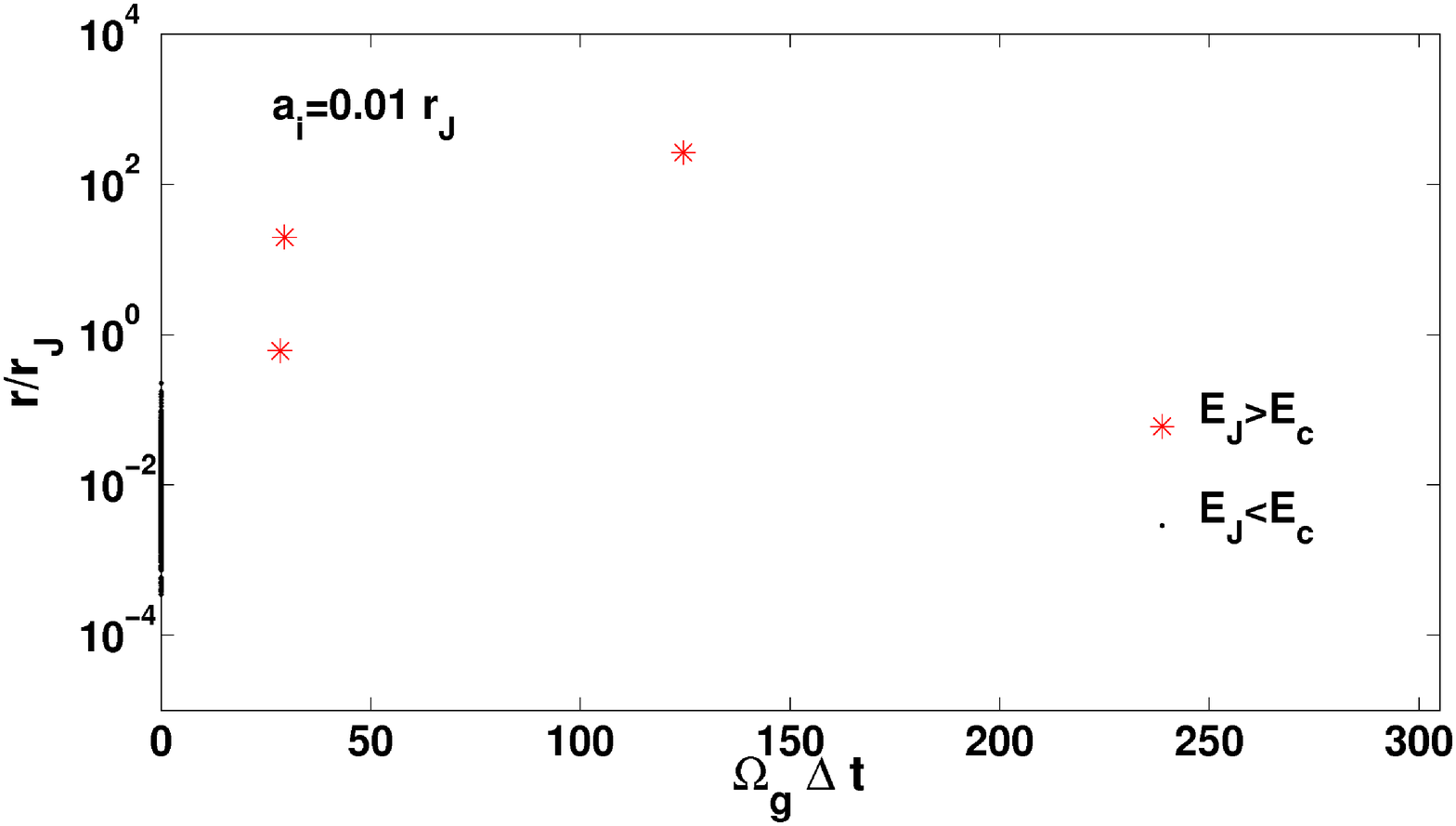}
\hfill
\includegraphics[width=0.49\hsize]{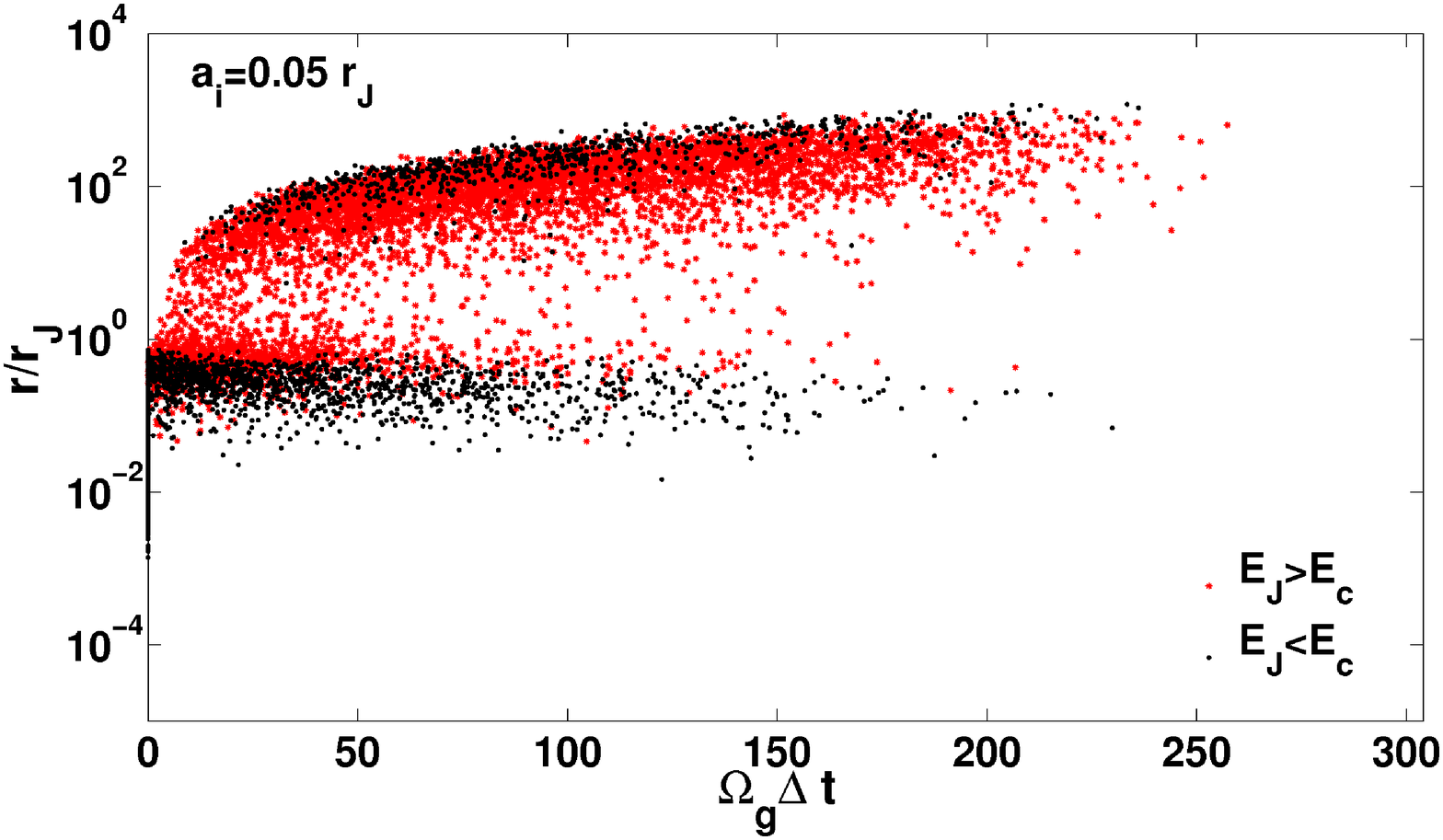} \\
\vspace{5mm}
\includegraphics[width=0.49\hsize]{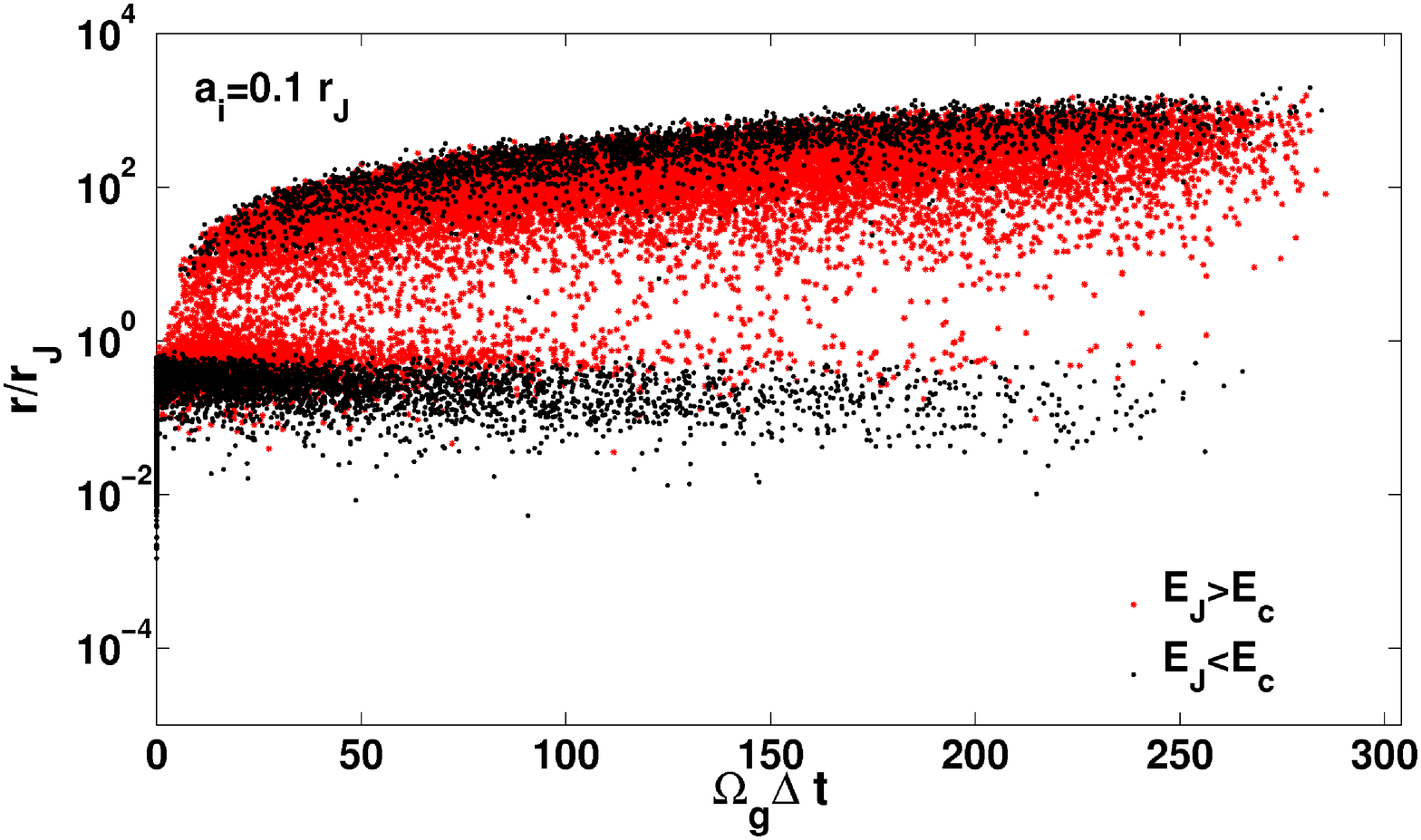}
\hfill
\includegraphics[width=0.49\hsize]{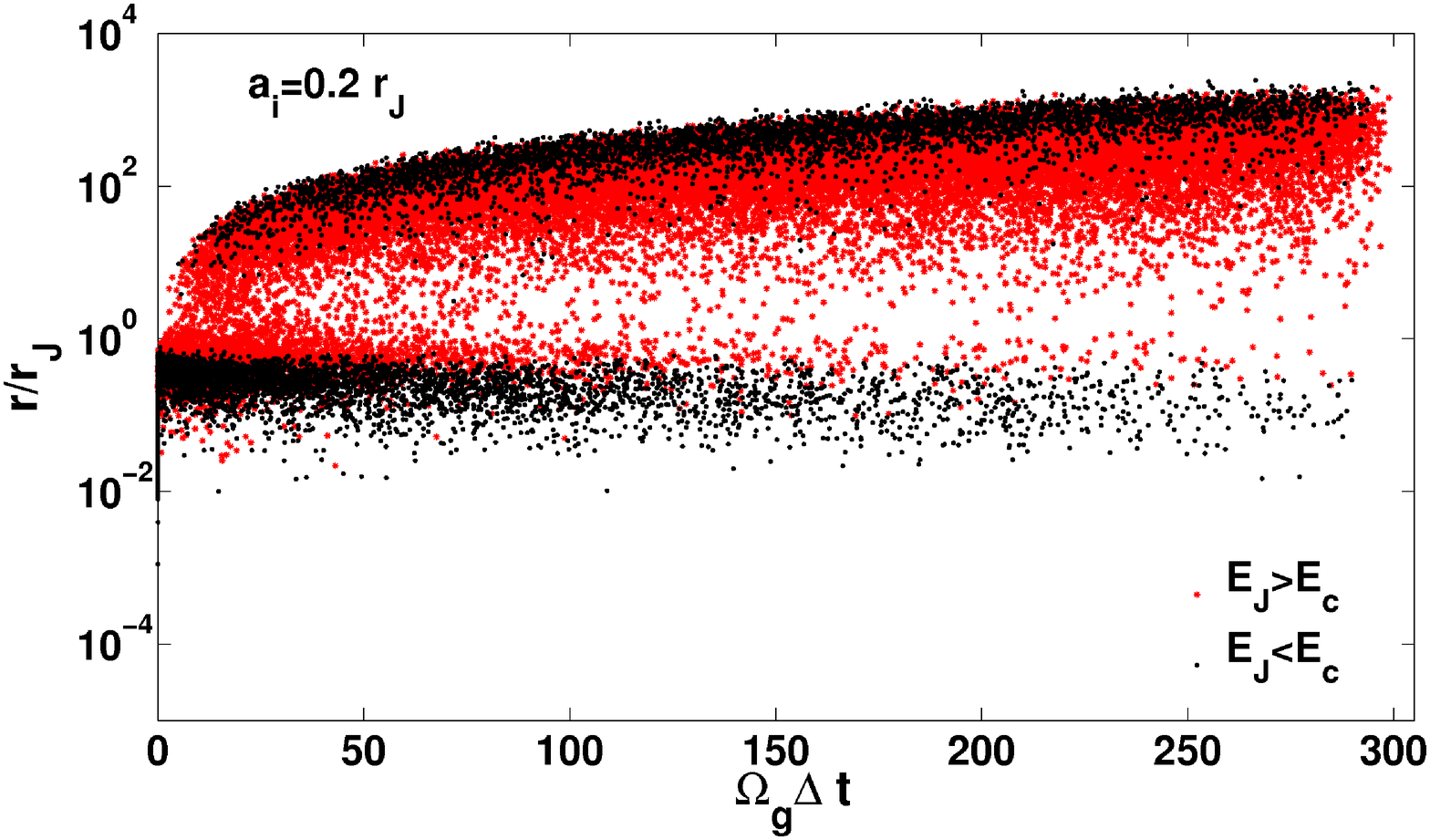} \\
\vspace{5mm}
\includegraphics[width=0.49\hsize]{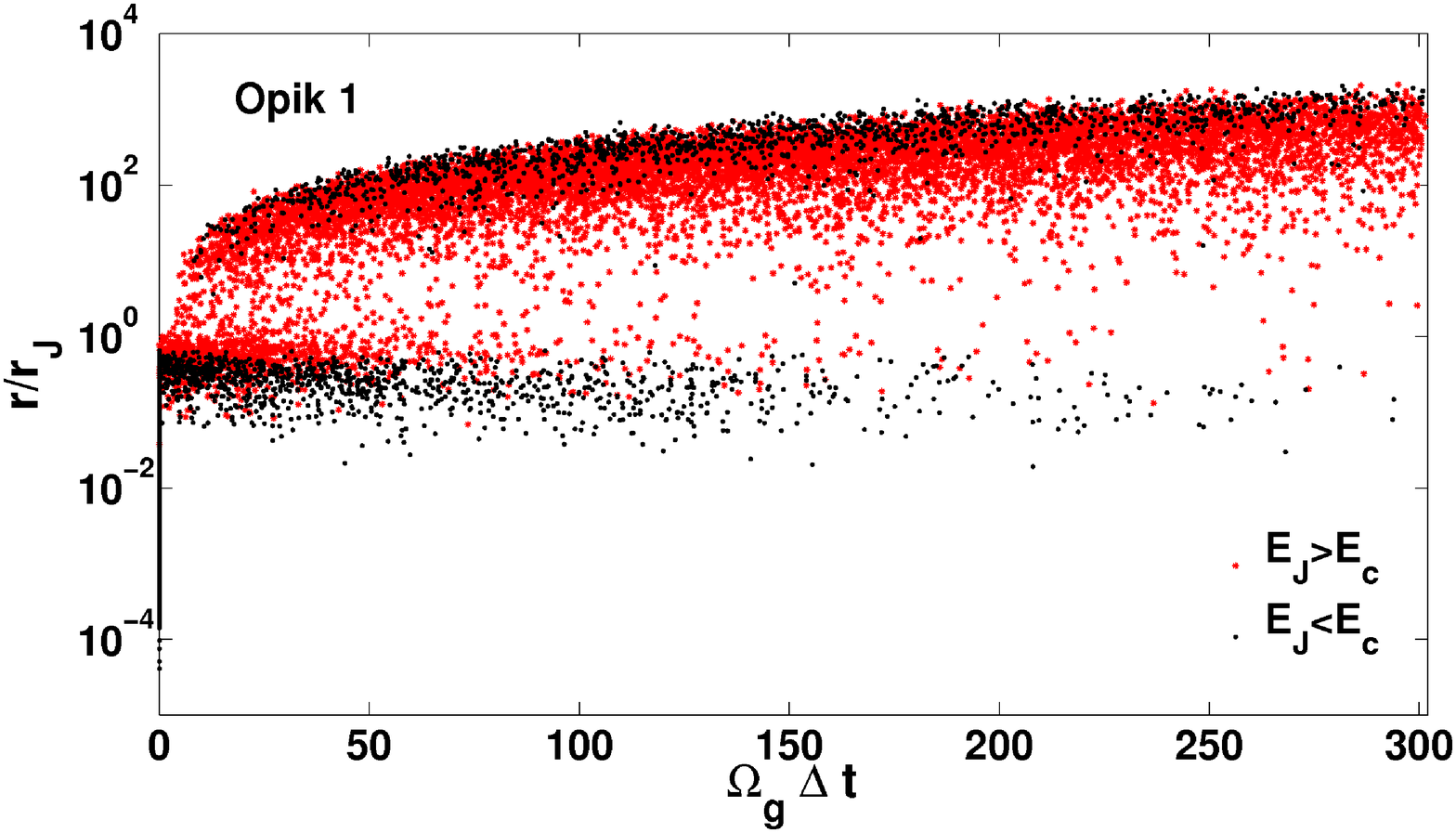}
\hfill
\includegraphics[width=0.49\hsize]{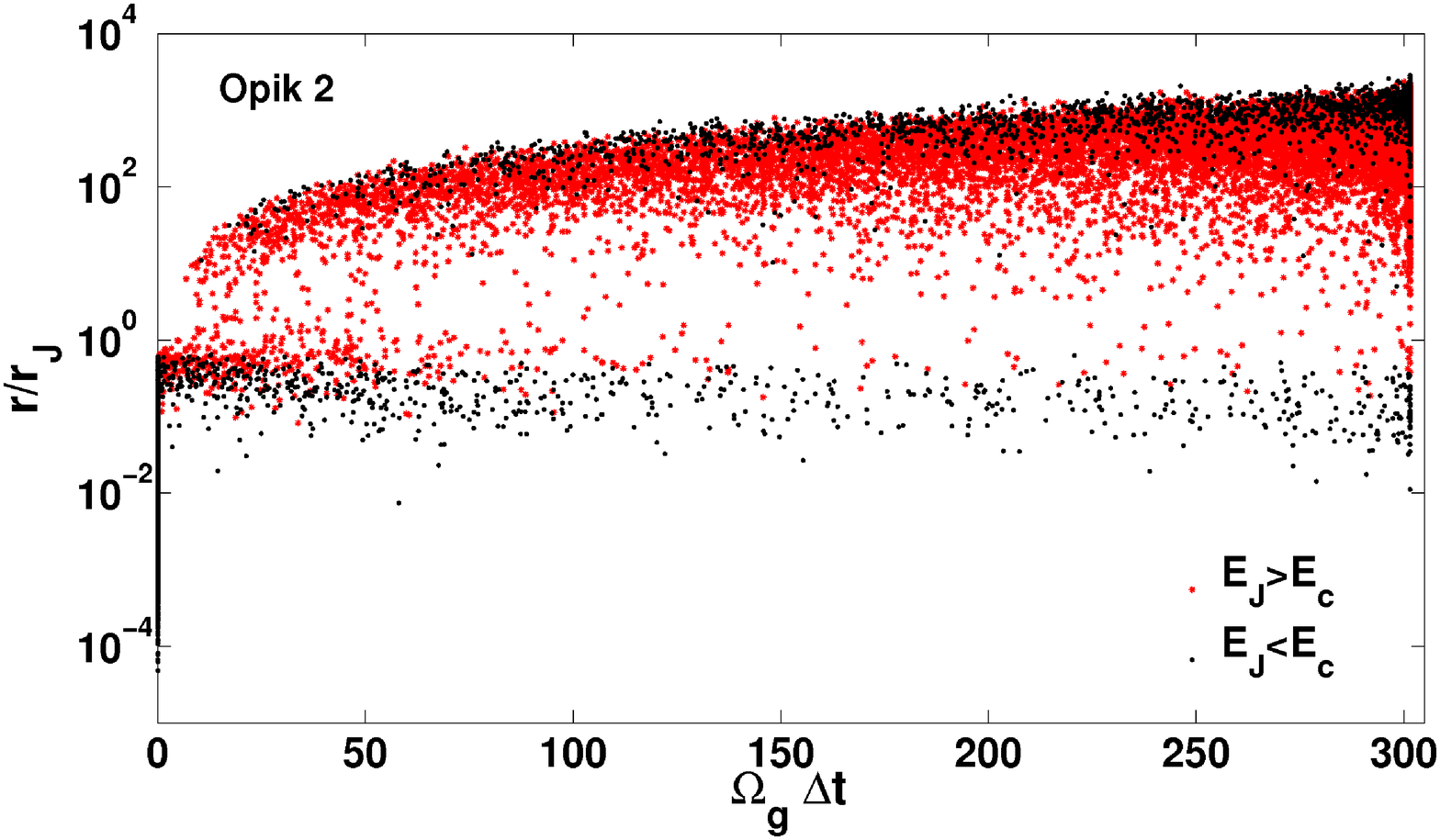} \\
\caption{Escape age $\Delta t$ versus separation of the binary
system $r$ at the end of the simulation. The initial conditions for
each panel are the same as in Figure \ref{phist}. The escape age is
the interval between the first instant when the Jacobi constant of
the binary system is larger than $E_c$ and the end of the
integration at $10\Gyr$. The black points are binary stars with
$E_J<E_c$ at the end of the simulation while the red points are
binaries with $E_J>E_c$. The black points with $\Delta t>0$ are
binary stars that have diffused back to $E_{J}<E_c$ even though
their Jacobi constants $E_J$ were larger than $E_c$ at the time they
escaped. The black points with $\Delta t=0$ are stars that never
escaped. There is a small concentration of points in the lower right
panel at $\Omega_g\Delta t\simeq 300$; these represent binaries that
were in escaped orbits at birth.}\label{initialtime}
\end{figure}

\begin{figure}
\includegraphics[width=0.49\hsize]{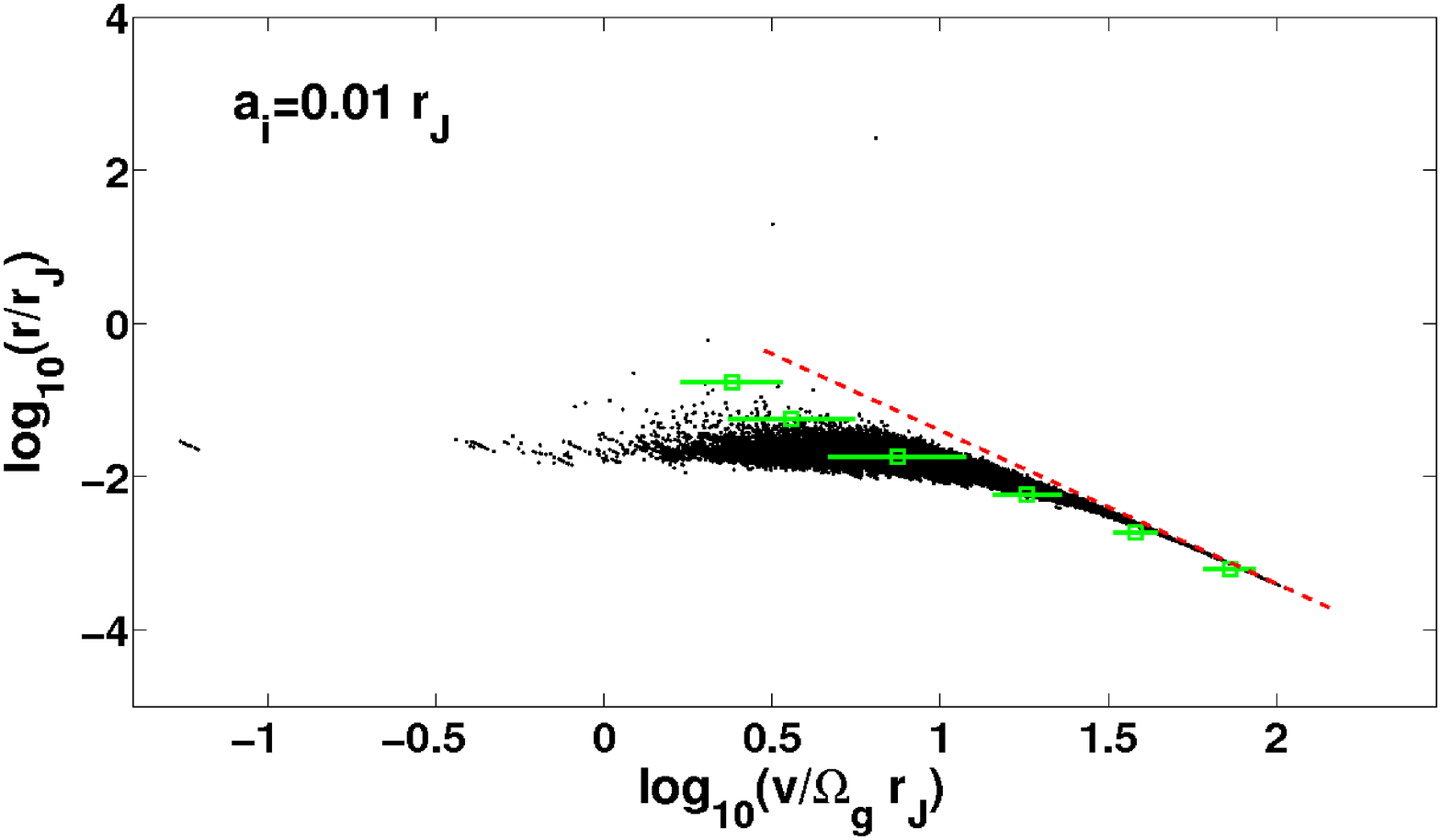}
\hfill
\includegraphics[width=0.49\hsize]{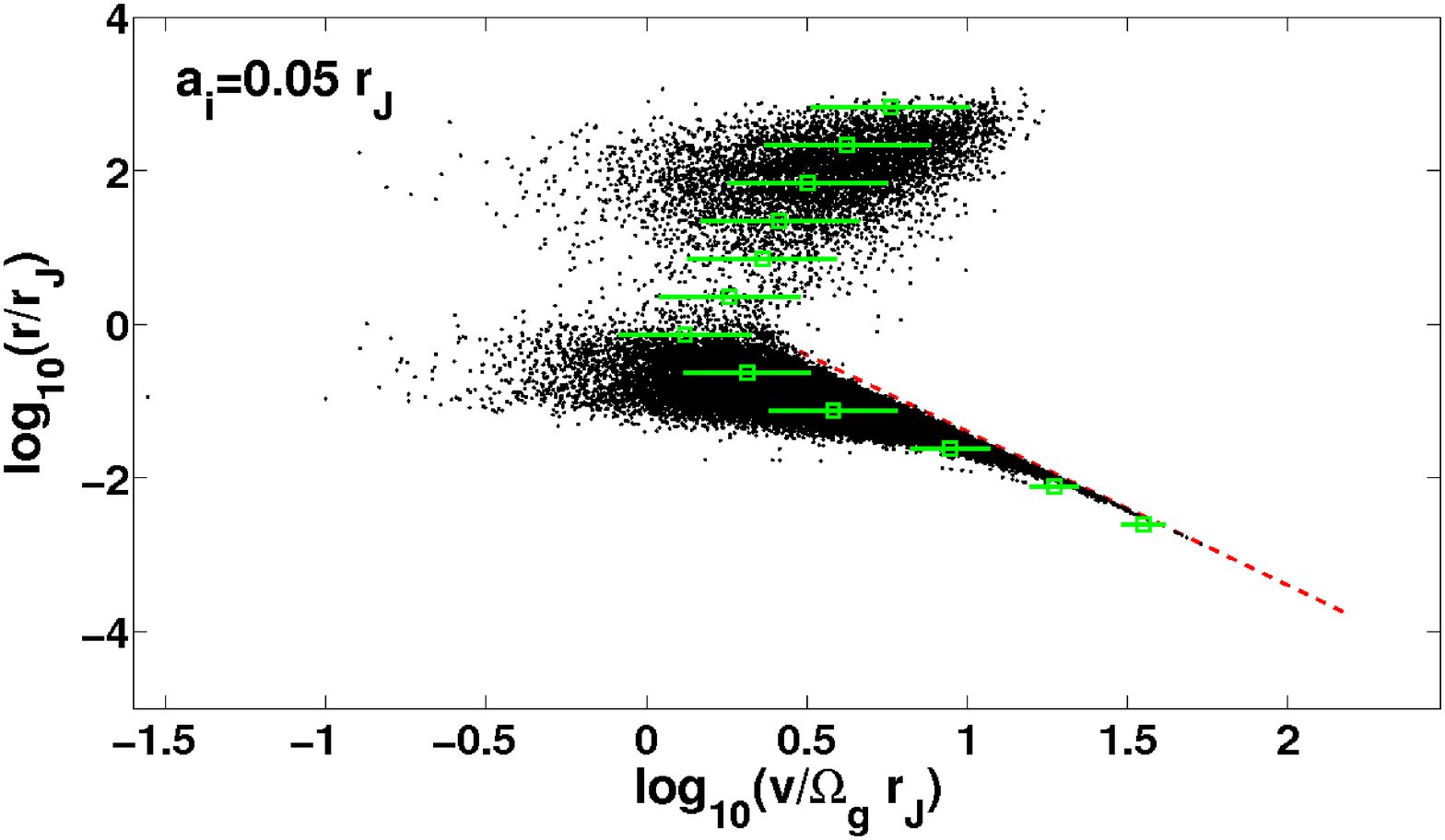} \\
\vspace{5mm}
\includegraphics[width=0.49\hsize]{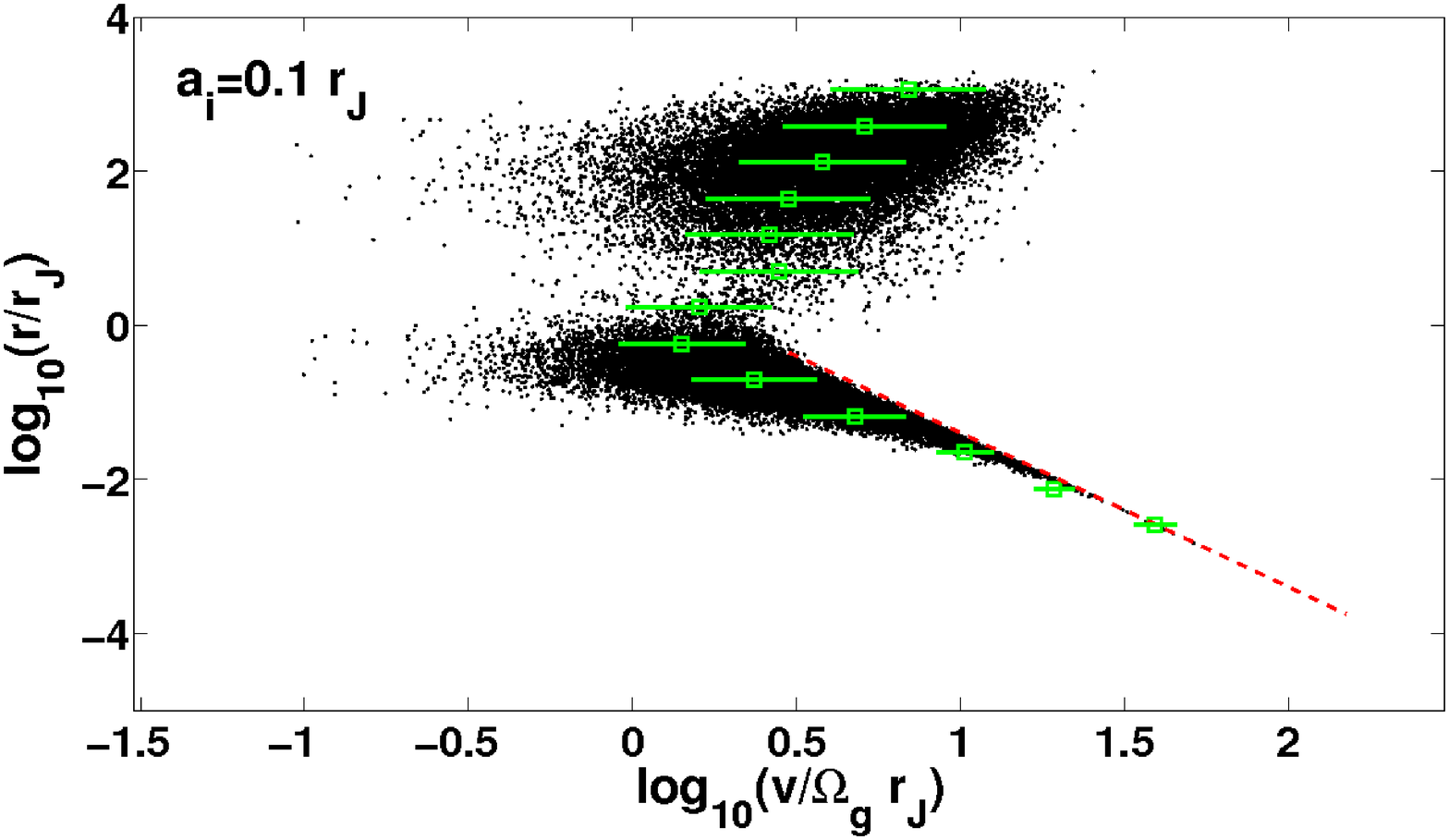}
\hfill
\includegraphics[width=0.49\hsize]{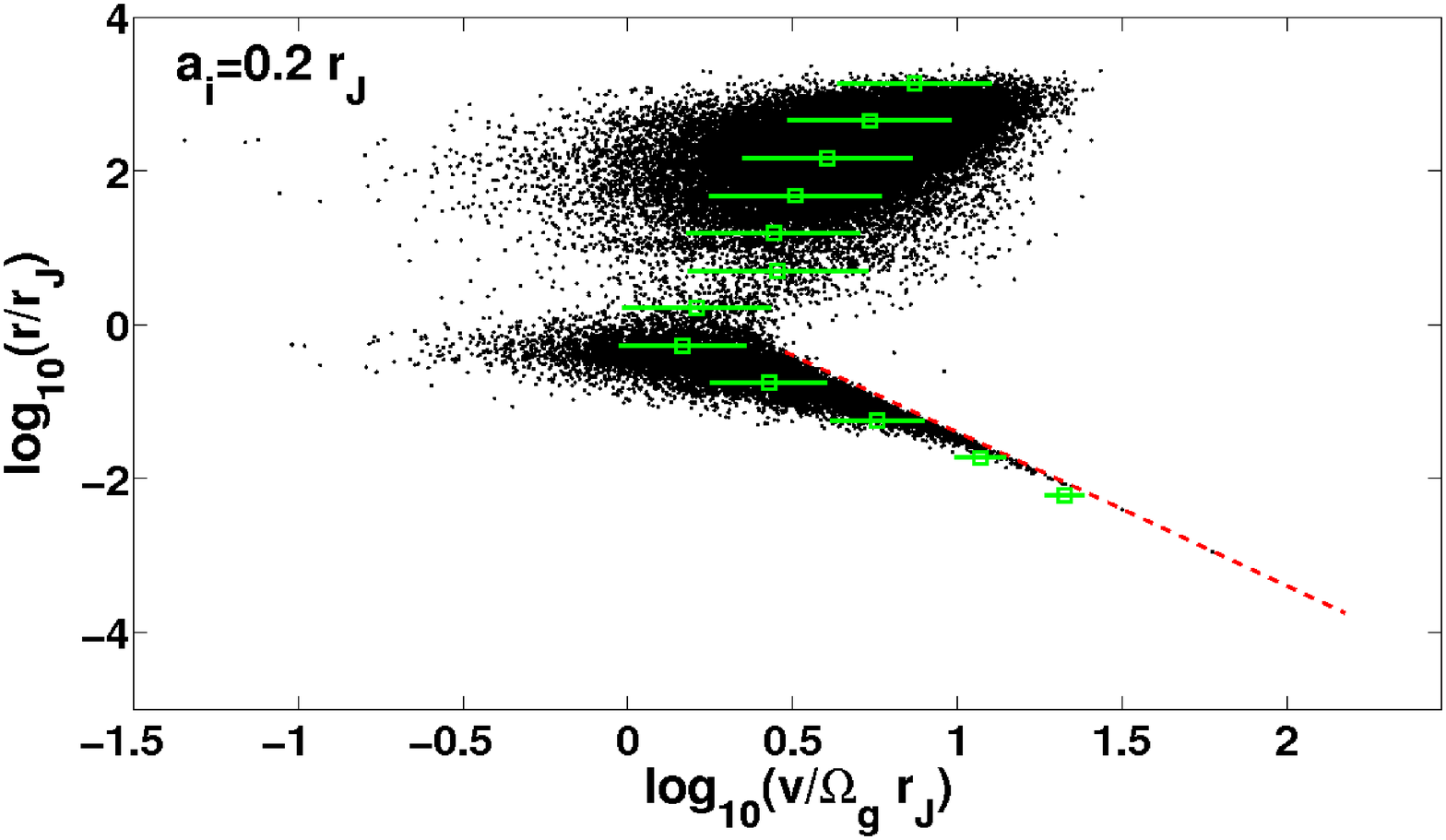} \\
\vspace{5mm}
\includegraphics[width=0.49\hsize]{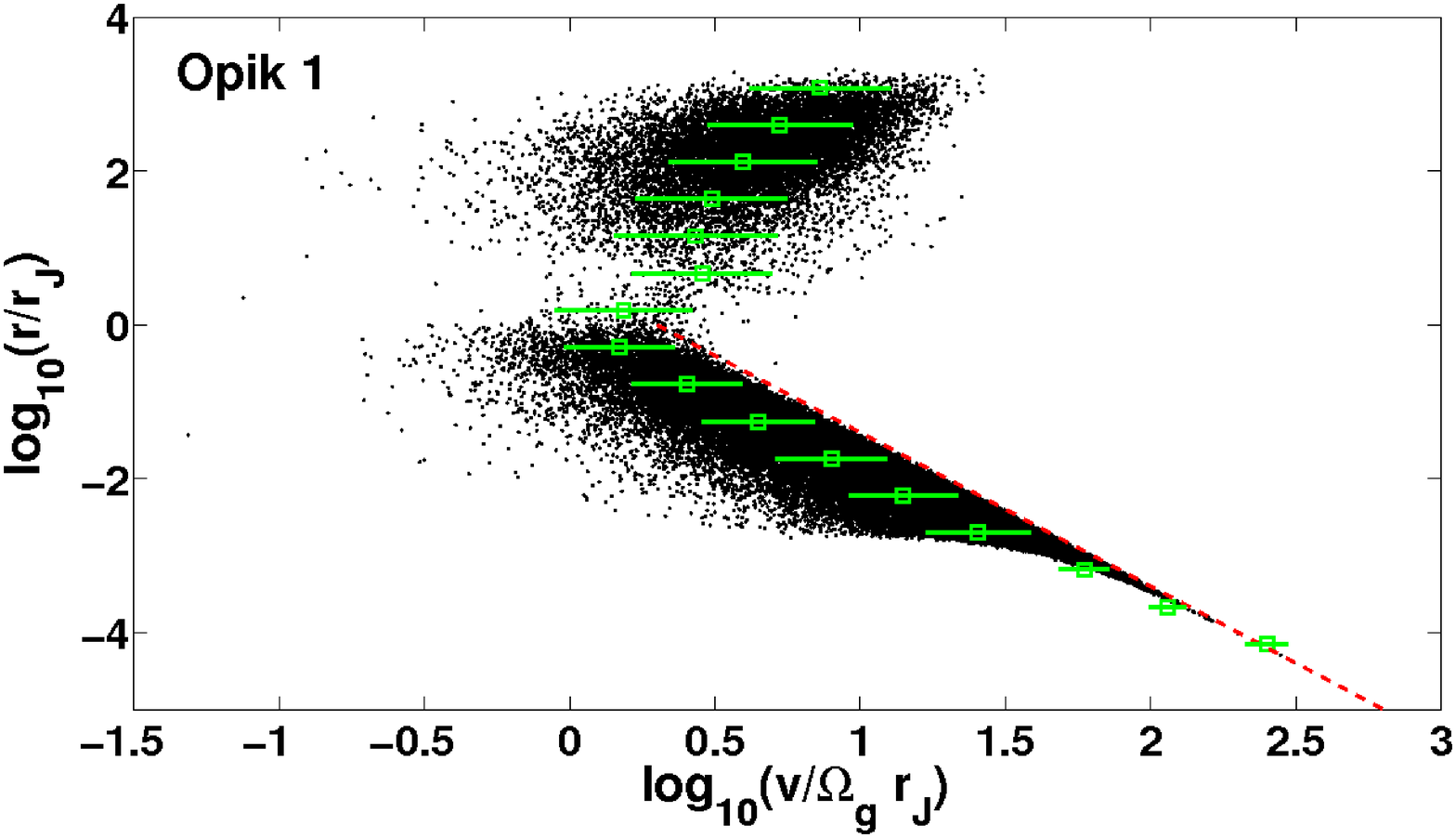}
\hfill
\includegraphics[width=0.49\hsize]{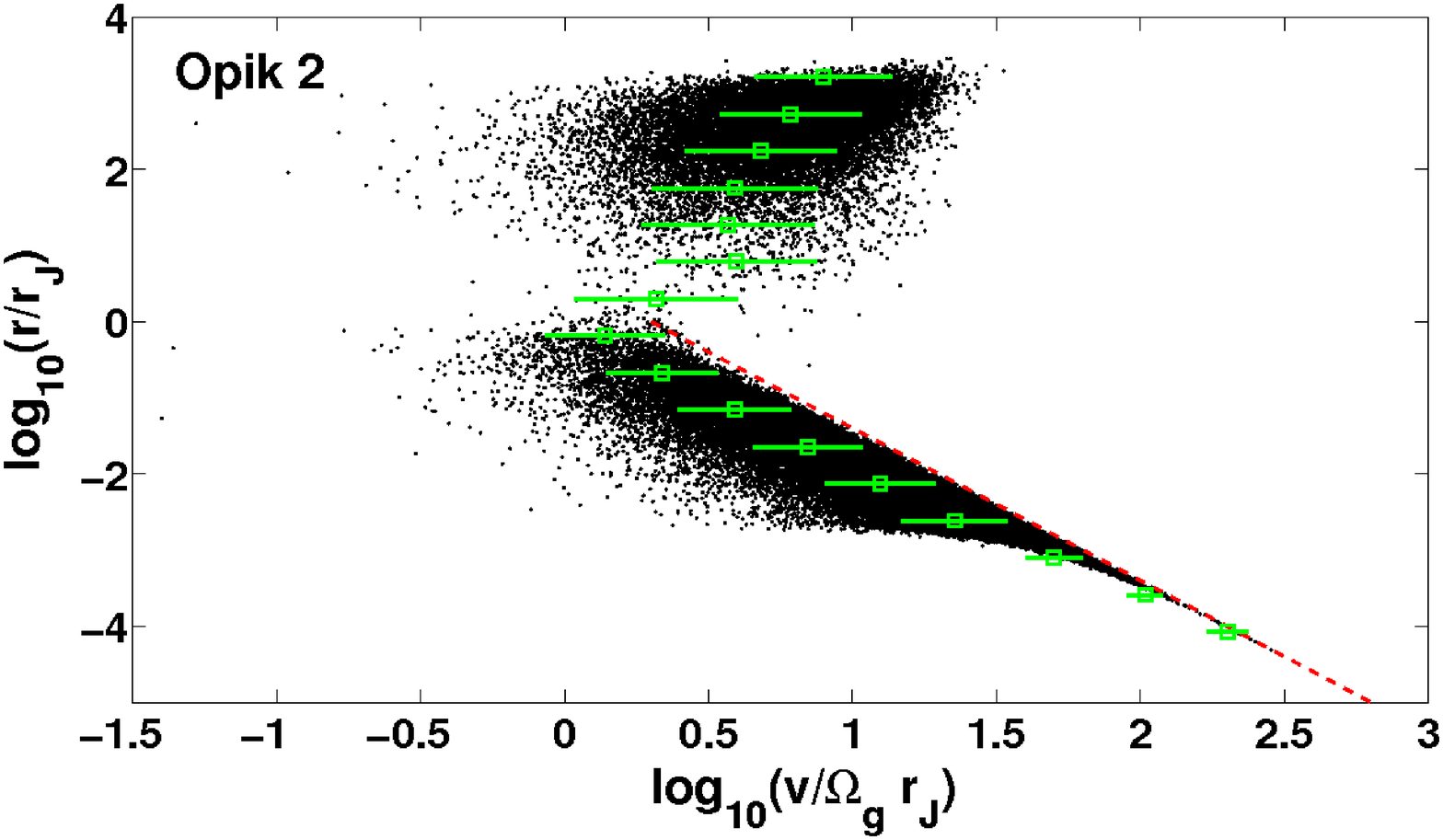} \\
\caption{Relative velocity and separation of the binaries at
$10\Gyr$. The
  initial conditions for each panel are the same as in Figure \ref{phist}. The
  red dotted line in each picture is the zero-energy line for Keplerian
  systems, $v^2/2=G(M_1+M_2)/r$ or
  $r/r_J=8(A_g/\Omega_g)(\Omega_gr_J/v)^2$.
  The binary stars with separation much smaller than $r_J$ follow the line
  very well. The green points with error bars
  show the mean and standard deviation in log velocity for various radius
  bins. The escaped binary stars mostly lie within a small velocity range
  ($|\Delta v|\lesssim 10\Omega_gr_J\simeq 0.5\kms$), which provides us with a
  good method to find escaped pairs. } \label{velocityradius}
\end{figure}

\subsection{Results from the numerical simulation}

\label{sec:results} The spatial distributions of the binary stars
after 10 Gyr are shown in Figure \ref{space}. In each panel the
relative position $\mathbf{r}=\mathbf{r}_1-\mathbf{r}_2$ is
projected onto the $x$--$y$ and $y$--$z$ plane. We can see tidal
tails along the $y$ direction (i.e., the direction of the binary's
Galactocentric orbit), which can extend to several thousands of
Jacobi radii.  As the initial semi-major axis $a_i$ increases, the
fraction of stars found in the tidal tails and the maximum extent of
the tidal tails both grow. In the case $a_i=0.01r_J$, only $11$
binaries of the original 50,000 have separation greater than
$10a_i=0.1r_J$ and only $2$ have separation greater than $r_J$. In
contrast, when $a_i=0.2r_J$, 70\% of the binaries have separation
greater than $10a_i=2r_J$ after 10 Gyr.

The distributions of projected separation (as viewed from a randomly
chosen position) for the binary stars in the six simulations are
shown in the histograms of Figure \ref{phist}. The blue histograms
show the full sample while the red histograms show binaries with
Jacobi constant $E_{J}>E_c$ at the end of the simulation.  The
figures also show the initial distributions of separations in green.
Binaries at large separations with $E_J<E_c$ must be in regions B or
C of Figure \ref{contour}, while binaries at large separations with
$E_J>E_c$ may be in any of regions B, C, D, or E.

Remarkably, rather than a cutoff in the distribution of binaries at
large separations, we see a local {\em minimum} in the density, at a
projected separation of about $5r_J$. (For binaries with initial
semi-major axis $a_0=0.01r_J$, the minimum is poorly defined as
there are only a few escapers.)  The distribution shows two peaks on
either side of the minimum; we call these the ``interior'' peak and
the ``exterior'' peak. Most of the stars in the interior peak are
bound (in the sense that they are found in region A of Figure
\ref{contour} and have $E_J<E_c$ so that in the absence of external
perturbations they must remain in region A forever). The fraction of
stars in the exterior peak grows as the initial semi-major axis
becomes larger or the age of the binaries grows. Note that the
minimum is present in plots like Figure \ref{phist} that show number
per unit logarithmic separation; plots of number per unit separation
are approximately flat between $5r_J$ and a few hundred $r_J$ but do
not show a minimum. Binary stars inside the interior peak in Figure
\ref{phist} roughly follow the initial distributions, shown in
green.

In the Appendix we describe a simple analytic model for the
distribution of separations that fits the simulations reasonably
well.

In Figure \ref{escaped}, we plot the distribution of separations of
the escaped binary stars only---stars that are outside region A of
Figure \ref{contour} or inside region A but with Jacobi constant
$E_J>E_c$ at 10 Gyr---for three different initial semi-major axes
$a_0=0.05r_J$, $0.1r_J$, $0.2r_J$.  As in Figure \ref{phist} there
is an ``exterior'' peak, and both the height of this peak and the
separation of the centroid of the peak grow with the initial
semi-major axis of the binaries. More surprising is that the
distribution of escaped stars in Figure \ref{escaped} also exhibits
an ``interior'' peak centered at $r\approx 0.5r_J$. The orbits of
the stars in this peak resemble those of the retrograde irregular
satellites of the giant planets, most of which are also formally
``escaped'' in the sense that their Jacobi constant $E_J>E_c$
\citep{hen70,st08}, but nevertheless can remain within $r_J$ for
very long times. Integrations for an additional $40\Gyr$, in which
kicks from passing stars were turned off, showed that the number of
stars in the interior peak declined with time only slowly, as
$t^{-0.1}$.

\begin{figure}
\includegraphics[width=\hsize]{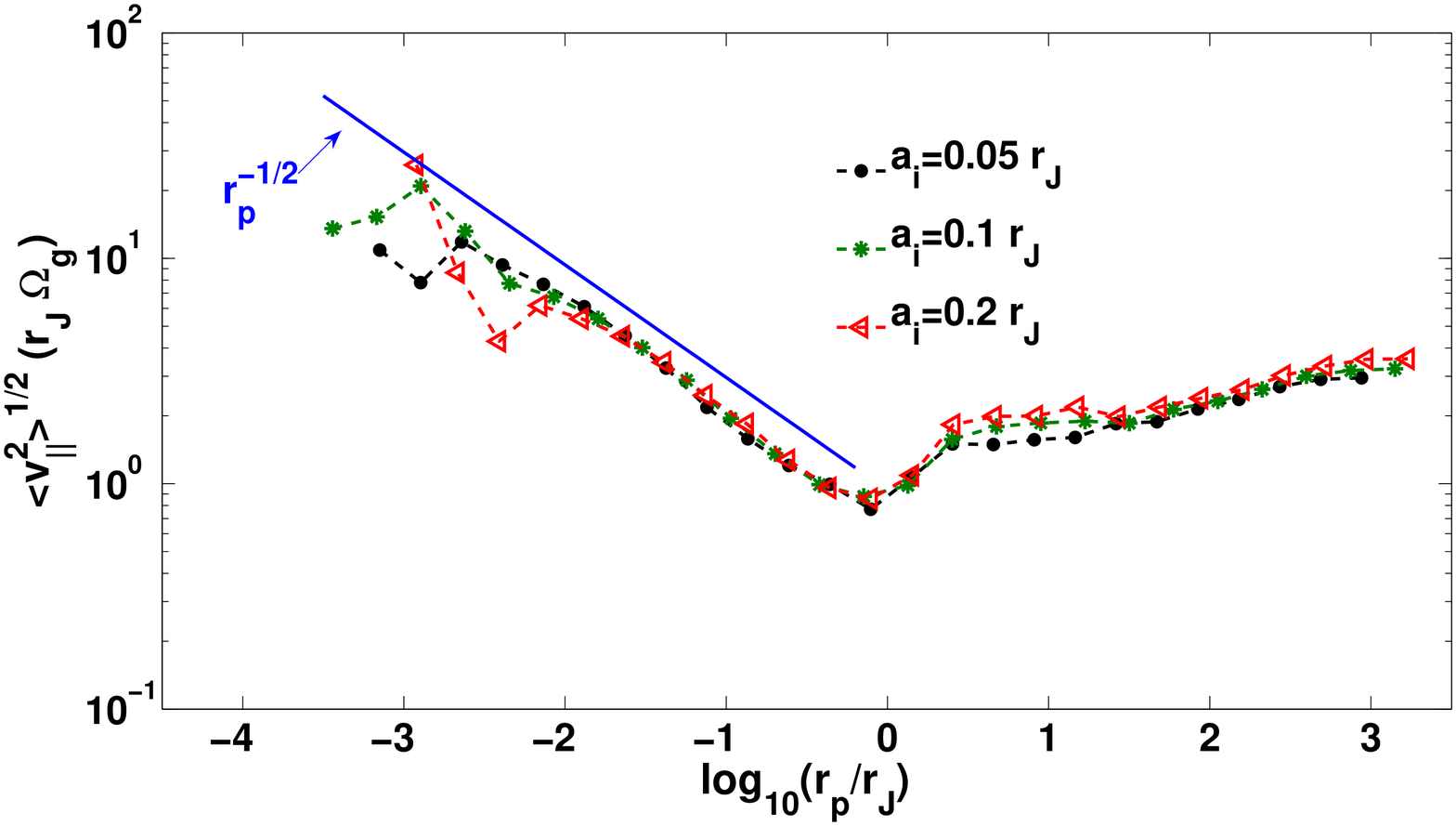}\\
\caption{RMS line-of-sight relative velocity of the binaries as a
function of projected separation, at the end of the simulations. The
horizontal axis is the projected separation normal
  to a randomly chosen line of sight, while the vertical axis is the RMS
  line-of-sight relative velocity in each separation bin. In Keplerian motion we expect
  $\langle v^2_{||}\rangle^{1/2}\propto r_p^{-1/2}$, shown by the straight
  line. The relation between the line-of-sight relative velocity and the
  projected separation deviates from the Keplerian relation for $r_p\gtrsim
  r_J$. }\label{averagevelocity}
\end{figure}

\begin{figure}
\includegraphics[width=\hsize]{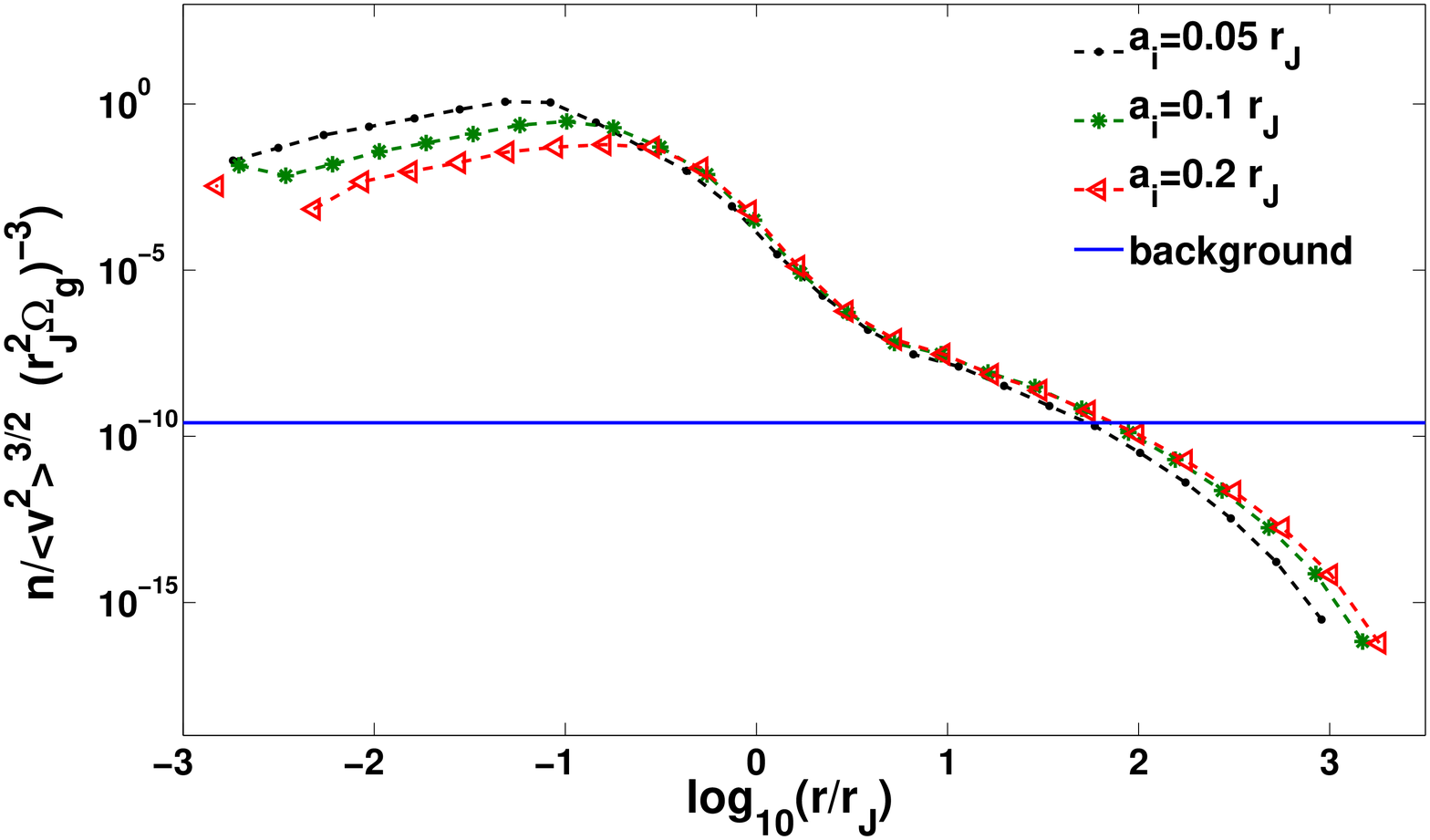}\\
\caption{Phase-space density of simulated binary stars compared with
  field stars in the solar neighborhood. The binary stars in each simulation
  at $10\Gyr$ are divided into different bins according to separation. The
  vertical axis is an indicative phase-space density, defined as the spatial
  density of binaries in each radius bin divided by $\langle
  v^2\rangle^{3/2}$, the cube of the RMS relative velocity. The density is
  normalized to the total number of binaries (50,000) in each simulation, and
  thus represents the phase-space density that would be observed in a catalog
  of $10^5/f$ stars where $f$ is the fraction of stars in the catalog
  that belong to wide binary systems. The
  blue horizontal line is the analogous indicative phase-space density for the
  field stars, computed as $\rho_0/\langle v^2\rangle^{3/2}$ where
  $\rho_{0}=0.68/r_J^3$ is given by equation (\ref{massdensity}) and $\langle
  v^2\rangle=3\sigma^2$ with $\sigma=40\kms$ as derived in
  \S\ref{valuesofparameters}. The phase-space density of binaries exceeds the
  density of field stars out to separations of $\sim 10^2r_J$.} \label{survey}
\end{figure}

We expect that the sooner the binary is disrupted---in the sense
that kicks from passing stars cause the Jacobi constant to random
walk to a value exceeding $E_c$---the larger the separation of the
binary system will be at the time $10\Gyr$. We label the interval
since the Jacobi constant of the binary first exceeded $E_c$ until
$10\Gyr$ (the ``escape age'') as $\Delta t$ (if the Jacobi constant
never exceeds $E_c$ we set $\Delta t=0$). The relation between the
separation at $10\Gyr$ and the escape age is shown in Figure
\ref{initialtime}. The red points have $E_{J}>E_c$ at $10\Gyr$ while
the black points have $E_J<E_c$. As noted earlier in Figures
\ref{phist} and \ref{escaped}, there is a gap around the separation
$5r_J$ in each panel. The black points with $\Delta t>0$ had
$E_J>E_c$ at some point in their history, but subsequent
perturbations kicked them back to $E_J<E_c$ (black points with
$r\gtrsim r_J$ must lie in regions B or C in Figure \ref{contour}).
The black points along the axis $\Delta t=0$ never escaped, i.e.,
$E_J<E_c$ for the entire integration. For the binary stars with
separation $r\gg r_J$, the general trend is that the sepration grows
with $\Delta t$. The upper envelope of the points in Figure
\ref{initialtime} is roughly $r\propto (\Delta t)^\alpha$ with
$\alpha=1.4$--1.5. This behavior has a simple physical explanation:
the relative velocity random walks due to stellar perturbations and
therefore grows as $v\propto (\Delta t)^{1/2}$, so the separation
grows as $r\sim v\Delta t \propto (\Delta t)^{1.5}$.

The relative velocity $v$ and separation $r$ of the binary systems
at the end of the simulation are shown in Figure
\ref{velocityradius}. The red dotted line is the zero-energy line
for Keplerian orbits, $v^2/2=G(M_1+M_2)/r$.  In the top four panels,
binary stars with separation $r$ much less than the initial
semi-major axis $a_i$ follow this line quite closely, since they are
generally found at $r\ll a_i$ only when they are near the pericenter
of near-parabolic orbits. In the bottom two panels, the binaries
follow the zero-energy line closely when $r\ll 0.001r_J$, the lower
cutoff to the semi-major axis range in the assumed initial \"Opik
distribution. As the separation increases, to $\sim r_J$, the
typical velocity decreases but the logarithmic spread in velocities
grows, as shown by the green error bars.

In Figure \ref{averagevelocity} we plot the relation between the RMS
line-of-sight relative velocity and projected separation, as seen
from an observer with a random orientation. Different initial
semi-major axes yield almost the same curve. When the separation is
$\lesssim r_J$, the relative velocity decreases with increasing
separation as $r_p^{-1/2}$, as one would expect for Keplerian
motion. When the separation is larger than $r_J$, the RMS relative
velocity increases with increasing separation. The minimum RMS
line-of-sight relative velocity is $\sim \Omega_gr_J$.

The maximum relative velocity for separations $\gg r_J$ is a few
times $\Omega_gr_J\simeq 0.05\kms$ (eq.\ \ref{eq:vjacdef}), two
orders of magnitude smaller than the typical relative velocity
between unrelated stars in the solar neighborhood. Because of this,
surveys that provide accurate velocity data have far greater ability
to identify binaries with $r\gg r_J$ than surveys with only
positions\footnote{D.\ Fabrycky points out that unbound
  pairs of asteroids, possibly formed by collisional disruption of large
  parent asteroids in the past, have been detected by similar techniques
  \citep{vn08}.}. To
illustrate this, in Figure \ref{survey} we plot the indicative
phase-space density (number density divided by $\langle
v^2\rangle^{3/2}$) of companions from our simulations, which contain
50,000 binary stars at birth. If a fraction $f$ of stars are found
in wide binaries, the total number of stars in a catalog that is
required to obtain 50,000 wide binaries is $1\times10^5/f$. The
horizontal line shows the analogous indicative phase-space density
of field stars in the solar neighborhood, $\rho_0/\langle
v^2\rangle^{3/2}$ where $\rho_{0}$ is given by equation
(\ref{massdensity}) and $\langle v^2\rangle=3\sigma^2$ with
$\sigma=40\kms$ as derived in \S\ref{valuesofparameters}. The
phase-space density of binaries exceeds the density of field stars
out to separations of $\sim 10^2r_J$ or well over $100\pc$. Thus a
statistical measurement of the distribution of binaries at $\sim
100\pc$ separation can be achieved by a survey such as GAIA that is
(i) large enough to contain $\gtrsim 10^5$ stars that were
originally in wide binaries; (ii) accurate enough that the errors in
distance and velocity are smaller than the separations and relative
velocities ($\sim 100\pc$ and $0.1$--$0.2\kms$).

\section{Discussion and conclusions}

\label{sec:disc}

We have studied the evolution and disruption of wide binary stars
under the gravitational influence of passing field stars. There have
been many treatments of this problem already (see the Introduction
for references) but most of these (i) ignore the Galactic tidal
field; (ii) define the binary to be ``disrupted'' when the Keplerian
energy becomes positive or when the separation exceeds the Jacobi or
tidal radius, and assume that the stars disappear instantaneously
once they are disrupted. The novel features of our treatment are
that we include the effects of the Galactic tidal field and follow
the evolution of the stars after they are disrupted.

Our simulations show that the usual treatment of binary disruption
is oversimplified. In particular,

\begin{itemize}

\item The number of binaries does not drop to zero when the separation exceeds
  the Jacobi radius $r_J$; rather there is a {\em minimum} in the density
  (number per unit log separation) at a few times $r_J$, almost
  independent of the initial semi-major axis distribution. Interior to this
  minimum there is a peak in the density due to the binaries that have not yet
  escaped, and exterior there is a peak due to binaries that are slowly
  drifting apart (Figure \ref{phist}).

\item Many binaries that have achieved escape energy (more precisely, that
  have Jacobi constants that exceed the critical value $E_c$ defined in eq.\
  \ref{eq:ecdef}) remain at separations less than the Jacobi radius for many
  Gyr, either because they are on stable orbits that do not escape to infinity
  or because subsequent perturbations from passing stars bring their Jacobi
  constant back below $E_c$ before they have time to escape (Figs.\
  \ref{escaped} and \ref{initialtime}).

\item Because the escaped binary components have small relative velocities,
  they contribute strongly to the phase-space correlation function in the
  solar neighborhood. Large astrometric surveys that can measure three-dimensional
  distances and velocities to sufficient accuracy ($\sim 100\pc$ and
  $0.1$--$0.2\kms$) can detect this correlation signal out to hundreds of
  parsecs.

\end{itemize}

These calculations could be improved in several ways. Our
simulations do not include perturbations from passing molecular
clouds, which are comparable to the perturbations from passing stars
at $a\sim0.1\pc$ within the uncertainties
\citep{ht85,Weinbergetal1987,mf01}.  Moreover the qualitative
effects of molecular clouds may be different because the impact
parameter of the most important cloud encounters is much larger than
the binary's Jacobi radius, whereas the most important stellar
encounters have impact parameters smaller than the Jacobi radius.
Molecular clouds have a much smaller scale height than old stars, so
the effects of passing clouds and stars may be disentangled
observationally by examining variations in the binary distribution
with the vertical amplitude of the center-of-mass motion of the
binaries \citep{sesaretal2008}.

The use of the central limit theorem to model stellar kicks as a
Gaussian distribution (eq.\ \ref{distribution}) is a plausible first
approximation but should eventually be replaced by a Monte Carlo
model of the kicks from individual passing stars. As described in
the discussion following equation (\ref{eq:follw}) the assumption
that there are many kicks per interval $\Delta t_p$ is not correct
at small semi-major axes. Also, for orbits near the critical Jacobi
constant $E_c$ there may be chaotic phenomena such as resonance
sticking that can only be modeled using the actual distribution of
velocity kicks \citep{hf00,heggie2001,ernst08}. Despite these
concerns, the tests we have carried out in
\S\ref{valuesofparameters} suggest that our results are not
sensitive to the specific value of $\Delta t_p$.

The distinction between evolution due to the Galactic tidal field
(\S\ref{sec:nokick}) and evolution due to impulsive kicks
(\S\ref{kick}) is artificial, since the same stars in the disk
contribute both the tidal field (apart from a contribution from dark
matter) and the kicks (apart from a contribution from molecular
clouds). The approximation that there is a static tidal field can be
misleading on timescales less than the collision time
(\ref{eq:tcolldef}); however, we do not believe that this
approximation has biased our results significantly. See \cite{ht86}
and \cite{cs09} for further discussions of this issue.

There is a large literature on tidal tails from star clusters
\citep[e.g.,][]{oden2001,belo06,gril06,kupperetal2008}. These differ
from the binary-star tails discussed here in several ways. Most
obviously, clusters contain many stars so the structure of the tail
from a single cluster can be mapped in great detail; in contrast,
the tail from a single binary contains only two stars so we must
combine many binaries to measure the tail properties. A second
difference is that the kicks to the orbits of stars in a cluster
arise from other cluster stars, and therefore cease once the star
escapes from the cluster, whereas the kicks to a binary arise from
passing stars and continue after disruption. The most important
consequence of this difference is that the length of a cluster tidal
tail grows $\propto t$, while a binary-star tail grows $\propto
t^{1.5}$.

Our results hold only for disk binary stars but it is
straightforward to repeat the calculation for halo binaries. These
are of particular interest because the semi-major axis distribution
of halo binaries can be used to constrain the mass distribution of
compact objects in the dark halo \citep{yooetal2004,quinnetal2009}.

\acknowledgments

We thank Dan Fabrycky, Mario Juri\'c, and Yue Shen for helpful
discussions. We also thank the referee, Winston Sweatman, for
comments that significantly improved the paper.  This research was
supported in part by NASA grant NNX08AH83G, and used computational
facilities supported by NSF grant AST-0216105.

\appendix

\section{Diffusion of the escaped binary stars}
\label{diffEBS}

Here we give an approximate analytic treatment of our results, by
solving for the evolution of binary stars with $r\ll r_J$ and $r\gg
r_J$ separately, then matching the two solutions at $r_J$.

The behavior of binary stars with separation much smaller than $r_J$
can be described by the diffusion approximation given in
\S\ref{diffusion}. The probability $p_e(\tau)d\tau$ for the binary
stars to escape in the time interval $(\tau,
\tau+d\tau)$\footnote{The initial time is set to be zero.} is
(derivative of eq.\ B19 in \citealt{Weinbergetal1987}, or from eq.\
\ref{escapetime} as $|E_1|\rightarrow0$)
\begin{equation}
p_e(\tau)=\frac{4}{3\sqrt{\pi}}\frac{h_0^{5/2}}{\tau^{7/2}}e^{-h_0/\tau}.
\label{eq:pedef}
\end{equation}

After the stars have escaped to $r\gg r_J$, the gravitational force
between the two stars is much smaller than the Galactic tidal force.
Then the relative motion in the absence of kicks is described by
equations (\ref{solutionrela}) with the right side set to zero.
Moreover their separations are dominated by drift along the
azimuthal or $y$ direction, as seen from Figure \ref{space}, so
$r\simeq y$. As the amplitude of the epicycle motion is small
compared to $y$, we have $y\approx y_g$, where $y_g$ is the position
of the guiding center of the relative motion (cf.\ the analogous
equations \ref{solutioncm} for the motion of the center of mass).
Therefore we must determine the equation that governs the evolution
of $y_g$ in the presence of kicks from passing stars.

The relation between the velocity of the guiding center $v_g=\dot
y_g$ and the relative position and velocity $(x,y,v_x,v_y)$ is
\begin{equation}
v_g=-\frac{A_g}{A_g-\Omega_g}(v_y+2\Omega_gx). \label{eq:vgv}
\end{equation}
This can be verified or derived from the epicycle equations for the
relative motion (the analogs of eqs.\ \ref{solutioncm} for the
center of mass epicycle motion) or from equations (8.101) and
(8.102) of \cite{binneytremaine2008}, which relate the orbital
parameters and the phase-space coordinates to the energy and angular
momentum in the epicycle approximation.

With equation (\ref{eq:vgv} and the impulse approximation for the
kick, the diffusion coefficient for $v_g$ is given by
\begin{equation}
D[(\Delta v_g)^2]=2\left(\frac{A_g}{A_g-\Omega_g}\right)^2D[(\Delta
v_y)^2]. \label{eq:vgdiff}
\end{equation}
The factor of two arises because kicks on both stars contribute to
the diffusion of $v_g$.

Let $f(t,y_g,v_g)dy_gdv_g$ be the probability that the escaped
binary stars lie in the interval $(y_g,y_g+dy_g)$ and
$(v_g,v_g+dv_g)$ at time $t$. Then the distribution function
$f(t,y_g,v_g)$ satisfies the simplified Fokker-Planck equation
\begin{equation}
\frac{\partial f}{\partial t}+v_g\frac{\partial f}{\partial
y_g}=\frac{1}{2}D[(\Delta v_g)^2]\frac{\partial^2f}{\partial v_g^2}.
\label{diffEBSequation}
\end{equation}
Here we have neglected the term $D[\Delta v_g]\partial f/\partial
v_g$ because $\partial f/\partial v_g$ is small compared to
$\partial^2f/\partial v_g^2$. The diffusion coefficient $D[(\Delta
v_y)^2]$ for either star is given by equation
(\ref{kickcoefficient}), which is now\footnote{The subscript $i$ for
the diffusion coefficients in equation (\ref{kickcoefficient}) is
omitted here.}
\begin{equation}
D[(\Delta v_y)^2]=\frac{v_y^2}{v^2}D[(\Delta
v_{||})^2]+\frac{v_x^2+v_z^2}{2v^2}D[(\Delta v_{\perp})^2].
\end{equation}
For binary stars with large separation, the velocity $v$ is much
smaller than $\sigma$. In this case, we have $D[(\Delta
v_{\perp})^2]=2D[(\Delta v_{||})^2]$ and thus $D[(\Delta
v_y)^2]=D[(\Delta v_{||})^2]$. Then from equations (\ref{diffcoef})
and (\ref{eq:vgdiff})
\begin{equation}
D[(\Delta v_g)^2]=\left(\frac{A_g}{A_g-\Omega_g}\right)^2
\frac{16\sqrt{2\pi}G^2\rho_{2}\ln\wedge}{3\sigma}.
\end{equation}
Note that the diffusion coefficient is independent of $y_g$ and
$v_g$ so we may label a constant $D_{g}\equiv D[(\Delta v_g)^2]$.
With the initial condition $f(0,y_g,v_g)=\delta(y_g)\delta(v_g)$ and
the boundary condition that $f\rightarrow 0$ when $y_g\rightarrow
\infty$ or $v_g\rightarrow \infty$, the solution to equation
(\ref{diffEBSequation}) is
\begin{equation}
f(t,y_g,v_g)=\frac{\sqrt{3}}{\pi
D_{g}t^2}\mbox{exp}\left[-\frac{6y_g^2}{D_{g}t^3}+
\frac{6y_gv_g}{D_{g}t^2}-\frac{2v_g^2}{D_{g}t}\right].
\end{equation}
The marginal probability distribution of $y_g$ can be gotten by
integration over $v_g$, which yields
\begin{equation}
f(t,y_g)=\int dv_g\,f(t,y_g,v_g)=\sqrt{\frac{3}{2\pi
D_{g}t^3}}\mbox{exp}\left[-\frac{3}{2}\frac{y_g^2}{D_{g}t^3}\right].
\end{equation}
Then at the final time $t_f=10$ Gyr, the probability that the binary
has separation $(y_g,y_g+dy_g)$] is
\begin{equation}
p_f(y_g)dy_g=dy_g\int_0^{t_f}p_e(t)f(t_f-t,y_g)dt, \label{distriyg}
\end{equation}
where $p_e(t)=p_e(\tau)d\tau/dt$ is given by (\ref{eq:pedef}).

\begin{figure}
\includegraphics[width=\hsize]{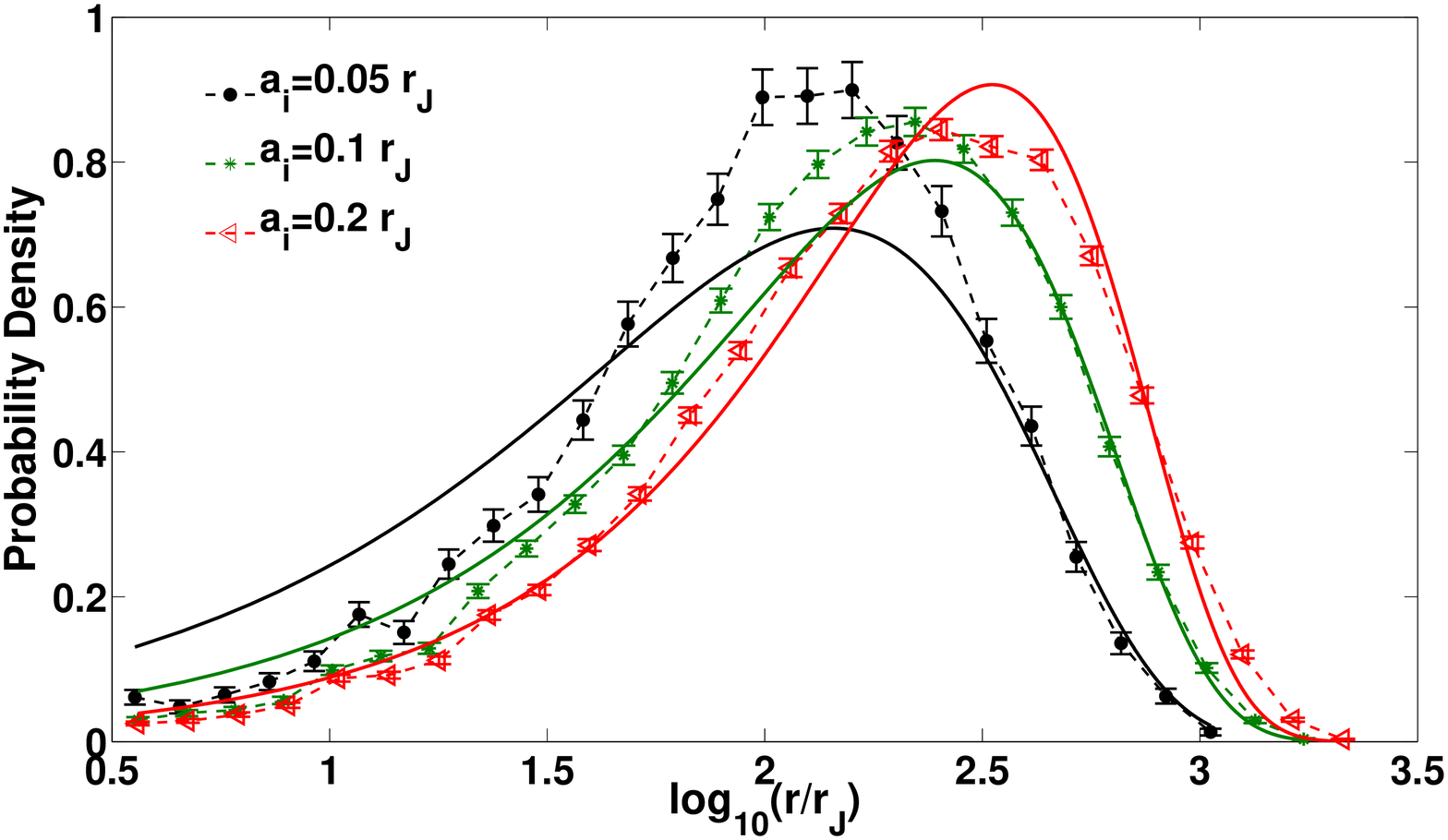}\\
\caption{Fit of the distribution of the binary stars outside the
minimum of
  the distribution in Figure \ref{escaped} ($r>3.16r_J$) to the analytic model
  described in the Appendix. All the lines are
  normalized to have unit area.  The
  dashed lines are data from the simulation while the solid lines are given by
  equation (\ref{distriyg}). The
  simulation with initial semi-major axis $a_0=0.01r_J$ is not shown here
  because of the small number of escaped stars. There are two reasons for the
  differences between the simulation and theoretical equation. The first is
  that we neglect the gravitational force within the binary system in equation
  (\ref{distriyg}), which is important near $r_J$. The second reason is that
  the maximum impact parameter $b_{\rm{max}}$ is actually not a constant
  during the simulation while in the theoretical equation we just choose a
  best fit value of $b_{\rm{max}}$, which is assumed to be a constant there.
  We can see the larger the initial semi-major axis is, the better our formula
  can fit the data.  }\label{fitescape}
\end{figure}

We compare the probability distribution (\ref{distriyg}) to the
escaped binary stars from our simulations in Figure \ref{fitescape}
(because the maximum impact parameter $b_{\rm{max}}$ is chosen to be
the half separation of the binary system at each kick time, which is
different at different times, we have to choose a ``mean''
$b_{\rm{max}}$ when we use equation (\ref{distriyg}) to fit the
simulation data).  We can see that the analytic treatment works
quite well at the largest separations, and works better if the
initial semi-major axis is larger. At small separations the fit is
less good, presumably because our approximation that the
gravitational force between the stars is negligible compared to the
tidal force is not accurate.

\end{document}